\documentclass[preprint,11pt]{elsarticle}
\usepackage{amsfonts}
\usepackage{epsfig,amsmath,latexsym,amssymb}
\usepackage{amscd}
\usepackage{multirow}
\usepackage{graphicx}
\usepackage{geometry}
\usepackage{lscape}
\usepackage{amsmath}
\usepackage{amssymb}
\usepackage{bm}
\usepackage{subfigure}

\providecommand{\U}[1]{\protect\rule{.1in}{.1in}}
\newcommand{\Remm}[1]{}
\newtheorem{theo}{Theorem}[section]

\newtheorem{cor}[theo]{Corollary}

\newtheorem{model ass}[theo]{Model Assumptions}

\numberwithin{equation}{section}

\begin{document}

\begin{frontmatter}

\title{Chain Ladder Method: Bayesian Bootstrap versus Classical Bootstrap}
\author{Gareth W.~Peters$^{1,2}$ \quad Mario V.~W\"{u}thrich$^{3}$ \ \ \ \
Pavel V.~Shevchenko$^{2}$}
\date{{\footnotesize {Working paper, version from \today }}}
\maketitle

\begin{abstract}
\noindent The intention of this paper is to estimate a Bayesian
distribution-free chain ladder (DFCL) model using approximate
Bayesian computation (ABC) methodology. We demonstrate how to
estimate quantities of interest in claims reserving and compare
the estimates to those obtained from classical and credibility
approaches. In this context, a novel numerical procedure utilising
Markov chain Monte Carlo (MCMC), ABC and a Bayesian bootstrap
procedure was developed in a truly distribution-free setting. The
ABC methodology arises because we work in a distribution-free
setting in which we make no parametric assumptions, meaning we can
not evaluate the likelihood point-wise or in this case simulate
directly from the likelihood model. The use of a bootstrap
procedure allows us to generate samples from the intractable
likelihood without the requirement of distributional assumptions,
this is crucial to the ABC framework. The developed methodology is
used to obtain the empirical distribution of the DFCL model
parameters and the predictive distribution of the outstanding loss
liabilities conditional on the observed claims. We then estimate
predictive Bayesian capital estimates, the Value at Risk (VaR) and
the mean square error of prediction (MSEP). The latter is compared
with the classical bootstrap and credibility methods.

\end{abstract}


\begin{center}
{\footnotesize {\ \textit{$^{1}$ UNSW Mathematics and Statistics
Department, Sydney, 2052, Australia; \\[0pt]
email: peterga@maths.unsw.edu.au \\[0pt]
$^{2}$ CSIRO  Mathematical and Information Sciences, Locked Bag
17, North Ryde, NSW, 1670, Australia \\[0pt]
$^{3}$ ETH Zurich, Department of Mathematics, CH-8092 Zurich,
Switzerland} } }
\end{center}

\end{frontmatter}

\pagebreak

\section{Motivation}

The distribution-free chain ladder model (DFCL) of Mack
\cite{Mack} is a popular model for stochastic claims reserving. In
this paper we use a time series formulation of the DFCL model
which allows for bootstrapping the claims reserves. An important
aspect of this model is that it can provide a justification for
the classical deterministic chain ladder (CL) algorithm which
originally was not founded on an underlying stochastic model.
Moreover, it allows for the study of prediction uncertainties.
Note that there are different stochastic models that lead to the
CL reserves (see for example W\"uthrich-Merz \cite{Wuthrich},
Section 3.2). In the present paper we use the DFCL formulation to
reproduce the CL reserves.

The paper presents a novel methodology for estimating a Bayesian
DFCL model utilising a framework of approximate Bayesian
computation (ABC) in a non-standard manner. A methodology
utilising Markov chain Monte Carlo (MCMC), ABC and a Bayesian
bootstrap procedure is developed in a distribution-free setting.
The ABC framework is required because we work in a
distribution-free setting in which we make no parametric
assumptions about the form of the likelihood. Effectively, the ABC
methodology allows us to overcome the fact that we cannot
evaluate the likelihood point-wise in the DFCL model. Typically,
ABC methodology circumvents likelihood evaluations by simulation
from the likelihood. However, in this case simulation from the
likelihood model is not directly available because no parametric
assumption is made. We combine ABC methodology with
bootstrap to overcome this additional complexity that the
DFCL model presents in the ABC framework. Then, by using an MCMC
numerical sampling algorithm combined with the novel version of
ABC that has the embedded bootstrap procedure, we are able to
obtain samples from the intractable posterior distribution of the
DFCL model parameters.

This allows us to utilise this methodology to obtain the Bayesian
posterior distribution of the DFCL model parameters empirically.
Then we demonstrate two approaches in which we can utilise the
posterior samples for the DFCL model parameters to obtain the
Bayesian predictive distribution of the claims. The first approach
involves using each posterior sample to numerically estimate the
full predictive claims distribution given the observed claims. The
alternative approach involves using the posterior samples for the
DFCL model parameters to form Bayesian point estimators. Then, conditional on these point estimators, we can obtain the Bayesian conditional predictive distribution for the claims. The second
approach will be relevant for comparisons with the classical and
credibility approaches. The first approach has the benefit that
it integrates out of the Bayesian predictive claims distribution
the parameter uncertainty associated with estimation of the DFCL
chain ladder parameters.

The paper then analyses the parameter estimates in the DFCL model,
the associated claims reserves and the mean square errors of
prediction (MSEP) from both the frequentist perspective and a
contrasting Bayesian view. In doing so we analyse CL point
estimators for parameters of the DFCL model, the resulting
estimated reserves and the associated MSEP from the classical
perspective. These include non-parametric bootstrap estimated
prediction errors which can be obtained via one of two possible
bootstrap procedures, conditional or unconditional. In this paper
we consider the process of conditional back propagation; see
\cite{Wuthrich} for in-depth discussion. These classical
frequentist estimators are then compared to Bayesian point
estimators. The Bayesian estimates considered are the maximum
\textit{a posteriori} (MAP) and the minimum mean square error
(MMSE) estimators. For comparison with the classical frequentist
reserve estimates, we also obtain the associated Bayesian estimated
reserves conditional upon the Bayesian point estimators.

In addition, since in the Bayesian setting we obtain samples from
the posterior for the parameters we use these along with the MSEP
obtained by the estimated Bayesian point estimators to obtain
associated posterior predictive intervals to be compared with the
classical bootstrap procedures. We then robustify the prediction
of reserves by Rao-Blackwellization, that is, we integrate out the
influence of the unknown variance parameters in the DFCL model.
Having done this, we analyse the resultant MSEP. This is again only
achievable since in the Bayesian setting we obtain samples from
the joint posterior for the CL factors and the variances.

To summarize our contribution, the novelty within this paper
involves the development and comparison of a new estimation
methodology to work with the Bayesian CL model for the DFCL model
which makes no parametric assumptions on the form of the
likelihood function; see also Gisler-W\"uthrich \cite{GW}. This is
unlike the works of Yao \cite{Yao} and Peters et al.
\cite{Peters1e} that assume explicit distributions in order to
construct the posterior distributions in the Bayesian context.
Instead we demonstrate how to work directly with the intractable
likelihood functions and the resulting intractable posterior
distribution, using novel ABC methodology. In this regard we
demonstrate that we do not need to make any parametric assumptions
to perform posterior inference, avoiding potentially poor model
assumptions made, as for example in the paper of Yao \cite{Yao}.

~ {\bf Outline of this paper.} The paper begins with a
presentation of the claims reserving problem and then presents the
model we shall consider. This is followed by the description of
the classical CL algorithm and the construction of a Bayesian
model that can be used to estimate the parameters of the model.
The Bayesian model is constructed in a distribution-free setting.
This is followed by a discussion on classical versus Bayesian
parameter estimators along with a bootstrap based procedure for
the estimation of the parameter uncertainty in the classical
setting. The next section presents the methodology of ABC coupled
with a novel bootstrap based sampling procedure which will allow
us to work directly with the distribution-free Bayesian model. We
then illustrate the developed algorithm on a synthetic data set
and the real data set, comparing performance to the classical
results and those obtained via credibility theory.

\section{Claims development triangle and DFCL model}

We briefly outline the claims development triangle structure we
utilise in the formulation of our models. Assume there is a
run-off triangle containing claims development data with the
structure given in Table 1.

{\footnotesize
\begin{table}[tbp]
\label{table:tab.1 Claims Development Triangle}
\begin{center}
{\scriptsize
\begin{tabular}{|c||rrrrrrrrrr|}
\hline
accident & \multicolumn{10}{c|}{development years $j$} \\
year $i$ & \quad0 \quad & \quad1 \quad & \quad\quad & \quad\quad &
\quad\quad & \enspace\dots\quad & \quad$j$\quad & \quad~ \quad &
\enspace\dots\quad & \quad$I$ \quad \\ \hline $0$ &
\multicolumn{10}{c|}{} \\ \cline{11-11}
$1$ & \multicolumn{9}{c|}{observed random variables $C_{i,j}\in\mathcal{D}%
_{I}$} & \multicolumn{1}{c|}{} \\ \cline{10-10} $\vdots$ &
\multicolumn{8}{c|}{} & \multicolumn{2}{c|}{} \\ \cline{9-9} &
\multicolumn{7}{c|}{} & \multicolumn{3}{c|}{} \\ \cline{8-8} $i$ &
\multicolumn{6}{c|}{} & \multicolumn{4}{c|}{} \\ \cline{7-7} &
\multicolumn{5}{c|}{} & \multicolumn{5}{c|}{} \\ \cline{6-6} &
\multicolumn{4}{c|}{} & \multicolumn{6}{c|}{} \\ \cline{5-5}
$\vdots$ & \multicolumn{3}{c|}{} & \multicolumn{7}{c|}{to be predicted ${C}%
_{i,j}\in\mathcal{D}_{I}^{c}$} \\ \cline{4-4} $I-1$ &
\multicolumn{2}{c|}{} & \multicolumn{8}{c|}{} \\ \cline{3-3} $I$ &
\multicolumn{1}{c|}{} & \multicolumn{9}{c|}{} \\ \hline
\end{tabular}
}
\end{center}
\caption{Claims development triangles.}
\end{table}
}

Assume that $C_{i,j}$ are cumulative claims with indices $i\in
\left\{ 0,\ldots ,I\right\}$ and $j\in \left\{ 0,\ldots
,J\right\}$, where $i$ denotes the accident year and $j$ denotes
the development year (cumulative claims can refer to payments,
claims incurred, etc). We make the simplifying assumption that the
number of accident years is equal to the number of observed
development periods, that is, $I=J$. At time $I$, we have observations
\begin{equation}
\mathcal{D}_{I}=\left\{ C_{i,j};~i+j\leq I\right\},
\end{equation}%
and for claims reserving at time $I$ we need to predict the future
claims
\begin{equation}
\mathcal{D}_{I}^{c}=\left\{ C_{i,j};~i+j>I,\;i\leq I,j\leq
J\right\} .
\end{equation}
Moreover, we define the set
$\mathcal{B}_{j}=\left\{ C_{i,k};~i+k\leq I, 0\leq k \leq
j\right\}$ for $j\in \{0,\ldots, I\}$, that is, $\mathcal{B}_{0}$ is the first column in Table 1.

\subsection{Classical chain ladder algorithm}

In the classical (deterministic) chain ladder algorithm there is
no underlying stochastic model. It is rather a recursive algorithm
that is used to estimate the claims reserves and which has proved
to give good practical results. It simply involves the following
recursive steps to predict unobserved cumulative claims in $
\mathcal{D}^c_I$. Set $\widehat{C}_{i,I-i}={C}_{i,I-i}$ and for
$j>I-i$
\begin{equation}
\widehat{C}_{i,j}=\widehat{C}_{i,j-1}\widehat{f}^{(CL)}_{j-1}\qquad
\text{ with CL
factor estimates }~\widehat{f}^{(CL)}_{j-1}=\frac{\sum\nolimits_{i=0}^{I-j}C_{i,j}}{%
\sum\nolimits_{i=0}^{I-j}C_{i,j-1}}.  \label{ChainLadderAlgorithm}
\end{equation}%
Since this is a deterministic algorithm it does not allow for
quantification of the uncertainty associated with the predicted
reserves. To analyse the associated uncertainty there are several
stochastic models that reproduce the CL reserves; for example
Mack's distribution-free chain ladder model \cite{Mack}, the
over-dispersed Poisson model (see England-Verrall \cite{EV}) or
the Bayesian chain ladder model (see Gisler-W\"uthrich \cite{GW}).
We use a time series formulation of the Bayesian chain ladder
model in order to use bootstrap methods and Bayesian inference.

\subsection{Bayesian DFCL model}

We use an additive time series version of the Bayes chain ladder
model (Model Assumptions 3.1 in Gisler-W\"uthrich \cite{GW}).

\begin{model ass}\label{model ass}~
\end{model ass}
\begin{enumerate}
\item We define the CL factors by $\mathbf{F}=\left(F_{0},\ldots,
F_{J-1}\right) $ and the standard deviation parameters by $\mathbf{\Xi }%
=\left( {\Xi }_{0},\ldots,{\Xi }_{J-1}\right) $. We assume
independence between all these parameters, i.e. the prior density
of $(\mathbf{F},\mathbf{\Xi })$ is given by
\begin{equation}
\pi(\bm{f},\bm{\sigma })= \prod_{j=0}^{J-1}
\pi(f_j)~\pi(\sigma_j),
\end{equation}
where $\pi(f_j)$ denotes the density of $F_j$ and $\pi(\sigma_j)$
denotes the density of $\Xi_j$.

\item Conditionally, given
$\bm{F}=\bm{f}=\left(f_{0},\ldots, f_{J-1}\right) $ and $\bm{\Xi
}=\bm{\sigma} =\left( \sigma_{0},\ldots,\sigma_{J-1}\right)$, we
have:
\begin{itemize}
\item Cumulative claims $C_{i,j}$ in different accident years $i$
are independent.
\item Cumulative claims satisfy the following time series
representation
\begin{equation}
C_{i,j+1}=f_{j}C_{i,j}+\sigma _{j}\sqrt{C_{i,j}}\varepsilon
_{i,j+1},
\end{equation}%
where conditionally, given $\mathcal{B}_0$, we have that the
residuals $\varepsilon _{i,j}$ are i.i.d.~satisfying
\begin{equation}
{E}\left[ \varepsilon _{i,j}|{\cal B}_0,\bm{F}, \bm{\Xi}\right]
=0~\text{ and }~\mathrm{Var}\left[ \varepsilon _{i,j}|{\cal
B}_0,\bm{F}, \bm{\Xi}\right] =1, \label{ResidualAssumptionsDFCL}
\end{equation}
and $P\left[\left.C_{i,j}>0\right|{\cal B}_0,\bm{F},
\bm{\Xi}\right]=1$ for all $i,j$.
\end{itemize}
\end{enumerate}

{\bf Remark.} Note that the assumptions on the residuals are
slightly involved in order to guarantee that cumulative claims
$C_{i,j}$ are positive $P$-a.s.

\begin{cor}\label{corind}
Under Model Assumptions \ref{model ass} we have that
conditionally, given $\mathcal{D}_I$, the random variables
$(F_0,\Xi_0),\ldots,(F_{J-1},\Xi_{J-1})$ are independent. 
Thus, we obtain the following posterior distribution for 
$(\bm{F},\bm{\Xi})$, given $\mathcal{D}_I$,
\begin{equation}
\pi \left( \bm{f},\bm{\sigma}|\mathcal{D}_{I}\right)
=\prod\limits_{j=0}^{J-1}\pi \left( f_{j},\sigma
_{j}|\mathcal{D}_{I}\right). \label{PosteriorIndependence}
\end{equation}%
\end{cor}
This result follows from Theorem 3.2 in Gisler-W\"uthrich \cite{GW};
from prior independence of the parameters; and the fact that
$C_{i,j+1}$ only depends on ${F}_j$, ${\Xi}_j$ and $C_{i,j}$
(Markov property). 
This has important implications for the ABC sampling algorithm
developed below.

In order to perform the Bayesian analysis we make explicit
assumptions on the prior distributions of $(\bm{F},\bm{\Xi})$.
\begin{model ass}~\label{model ass2}\end{model ass}
In addition to Model Assumptions \ref{model ass} we assume that
the prior model for all parameters $j\in \left\{ 0,\ldots
,J-1\right\} $ is given by:
\begin{itemize}
\item $F_{j}\sim \Gamma \left( \alpha_j ,\beta_j
\right) $, where $\Gamma \left( \alpha_j ,\beta_j \right) $ is a
gamma distribution with mean ${E}\left[ F_{j}\right] =\alpha_j \beta_j =%
\widehat{f}_{j}^{(CL)}$ (see \eqref{ChainLadderAlgorithm}) and
large variance to have diffuse priors.

\item The variances ${\Xi }^2_{j}\sim IG\left( a_j,b_j\right) $,
where $%
IG\left( a_j,b_j\right) $ is an inverse gamma distribution with mean ${E}%
\left[ {\Xi }^2_{j}\right] =b_j/(a_j-1)=\widehat{\sigma }%
_{j}^{2(CL) }$ (see \eqref{CCL_Var_DFCL} below) and large
variance.
\end{itemize}

{\bf Remarks}

\begin{enumerate}
\item The likelihood model is intractable, meaning that no density
can be written down analytically in the DFCL model. In formulating
the Bayesian model we have only made distributional assumptions on
the priors for the parameters $(\bm{F},\bm{\Xi})$ but not on the
observable cumulative claims $C_{i,j}$. Though we make
distributional assumptions for the priors, the model is
distribution-free because no distributional assumptions on the
cumulative claims are made. As a result of only making assumptions
on the priors, a standard Bayesian analysis using analytic
posterior distributions cannot be performed. One way out of this
dilemma would be to re-formulate the Bayesian model by making
distributional assumptions (for example, this is done in Yao \cite{Yao})
but then the model is no longer distribution-free. Another
approach would be to use credibility methods (see
Gisler-W\"uthrich \cite{GW}) but this only gives statements for
the first two moments. In the present set up we develop ABC
methods that allow for a full distributional answer for the
posterior distributions without making explicit distributional
assumptions for the cumulative claims $C_{i,j}$.

\item Our priors are chosen as diffuse priors with large variances.
This again highlights the differences between specification of the
prior distributions and making distributional assumptions for the
actual likelihood model, these are mutually exclusive ideas.

\item We select the priors to ensure that we maintain several
relevant aspects of the DFCL model. In particular, it is important
to utilise priors that enforce the strict positivity of the
parameters $f_j,\sigma_j >0$. We note here that the parametric
Bayesian model developed in Yao \cite{Yao} failed in this aspect
when it came to prior specification. Therefore we develop an
alternative prior structure that satisfies these required
properties of the DFCL model.

\end{enumerate}

\section{DFCL model parameter estimators}
This section considers both classical and Bayesian estimators for the chain ladder framework, including both the chain ladder factors and the variance parameters. 
\subsection{Classical}
In the classical CL method, the CL factors are estimated by
$\widehat{f}_j^{(CL)}$ given in \eqref{ChainLadderAlgorithm}. The
variance parameters are estimated by
\begin{eqnarray}
\widehat{\sigma }_{j}^{2(CL) } &=&
\frac{1}{I-j-1}\sum\nolimits_{i=0}^{I-j-1}C_{i,j}\left( \frac{C_{i,j+1}}{%
C_{i,j}}-\widehat{f}_{j}^{(CL) }\right) ^{2};
\label{CCL_Var_DFCL}
\end{eqnarray}
see (3.4) in W\"{u}thrich-Merz \cite{Wuthrich}.

Note that this estimator is only well-defined for $j<I-1$. There
is a vast literature and discussion on the estimation of tail
parameters. We do not enter this discussion here but we simply
choose the estimator given in Mack \cite{Mack} for the last
variance parameter which is defined by
\begin{equation}
\widehat{\sigma }_{J-1}^{2(CL) } =
\min \left\{ \frac{\widehat{\sigma }_{J-2}^{4(CL) }}{\widehat{%
\sigma }_{J-3}^{2(CL) }},\widehat{\sigma }_{J-3}^{2( CL)
},\widehat{\sigma }_{J-2}^{2(CL) }\right\}.
\end{equation}

\subsection{Bayesian}
In a Bayesian inference context one calculates the posterior
distribution of the parameters, given $\mathcal{D}_I$. As in
\eqref{PosteriorIndependence} we denote this posterior by $\pi
\left( \bm{f},\bm{\sigma}|\mathcal{D}_{I}\right)$. Since the
MCMC-ABC bootstrap procedure will allow us to obtain samples from
the posterior distribution of the Bayesian DFCL model presented,
we can now consider estimating CL point estimators using these
samples.

There are two commonly used point estimators in Bayesian analysis
that correspond to the posterior mode (MAP) and the posterior mean
(MMSE), respectively:
\begin{eqnarray}
\left(\widehat{f}_{j}^{(MAP) }, \widehat{\sigma }_{j}^{(MAP) }\right) &=&{\arg \max}_{f_j,\sigma_j}  ~\pi \left( f_{j}, \sigma_j|\mathcal{D}_{I}\right),  \label{MAP_f} 
\end{eqnarray}%
and
\begin{eqnarray}
\widehat{f}_{j}^{( MMSE) } &=&\int f_{j}~\pi \left(
f_{j}|\mathcal{D}_{I}\right) df_{j}= E\left[\left. F_j\right|\mathcal{D}_I \right],  \label{MMSE_f} \\
\widehat{\sigma }_{j}^{( MMSE) } &=&\int ~\sigma _{j}~\pi \left(
\sigma _{j}|\mathcal{D}_{I}\right) d\sigma _{j}=E\left[\left.
\Xi_j\right|\mathcal{D}_I \right]. \label{MMSE_var}
\end{eqnarray}
In the case in which $f_j$ is not independent of $\sigma_j$, the MAP estimators obtained through joint maximization are optimal. However, in practice one often works with marginal estimators for simplicity. Additionally, note that for diffuse priors we find (see Corollary 5.1 in
Gisler-W\"uthrich \cite{GW})
\begin{equation}
\widehat{f}_{j}^{( MMSE) } \approx \widehat{f}_{j}^{(CL) }.
\end{equation}
Hence, using Corollary \ref{corind}, we obtain the approximation
\begin{eqnarray}
E\left[\left. C_{i,J}\right|\mathcal{D}_I \right] &=&E\left[\left.
E\left[\left. C_{i,J}\right|\mathcal{D}_I,\bm{F}, \bm{\Xi}
\right]\right|\mathcal{D}_I \right]=C_{i,I-i}~E\left[\left.
\prod_{j=I-i}^{J-1}F_j\right|\mathcal{D}_I \right]
\nonumber\\
\label{BayesPredictor} &=&C_{i,I-i}~
\prod_{j=I-i}^{J-1}E\left[\left. F_j\right|\mathcal{D}_I \right]
~=~C_{i,I-i}~ \prod_{j=I-i}^{J-1}\widehat{f}_{j}^{( MMSE) }
\\\nonumber
&\approx &C_{i,I-i}~ \prod_{j=I-i}^{J-1}\widehat{f}_{j}^{(CL) }
~=~\widehat{C}_{i,J},
\end{eqnarray}
where on the last line we have an equality if the
diffusivity of the priors $\pi(f_j)$ tends to infinity. This is
exactly the argument why the Bayesian CL  model can be used to
justify the CL predictors; see Gisler-W\"uthrich \cite{GW}.

\subsection{Full predictive distribution and VaR}
In addition, the posterior samples for the DFCL model parameters,
obtained via the MCMC-ABC bootstrap procedure, will allow us to
obtain the predictive distribution of the claims in two ways. The
first is the full predictive distribution of the claims obtained
after integrating out the posterior uncertainty associated with
the Bayesian DFCL model parameters to empirically estimate
\begin{equation}
\pi\left(\mathcal{D}^{c}_I|\mathcal{D}_I\right) = \int \int
\pi\left(\mathcal{D}^{c}_I|\bm{f},\bm{\sigma}\right)
\pi\left(\bm{f},\bm{\sigma}|\mathcal{D}_I\right) d\bm{f}
d\bm{\sigma}.
\end{equation}
In practice, this numerical procedure involves taking each
posterior sample for the DFCL model parameters and obtaining an
estimate of the predicted claims.

The second approach involves using one of the Bayesian point
estimators for the parameters such as the MMSE to obtain
$\pi\left(\mathcal{D}^{c}_I|\widehat{\bm{f}}^{MMSE},\widehat{\bm{\sigma}}^{MMSE}\right)$. 
Alternatively, one may consider a Rao-Blackwellised version of the Bayesian predictive
distribution of claims involving 
$$\pi\left(\mathcal{D}^{c}_I|\widehat{\bm{f}}^{MMSE},\mathcal{D}_I\right) = \int
\pi\left(\mathcal{D}^{c}_I|\widehat{\bm{f}}^{MMSE},\bm{\sigma}\right)\pi\left(\bm{\sigma}|\widehat{\bm{f}}^{MMSE},\mathcal{D}_I\right) d\bm{\sigma}$$
having numerically integrated out the Bayesian posterior
uncertainty associated with the DFCL variance parameters. Such methods are typically known as empirical Bayesian approaches.

These results can then be applied to estimate any risk measures.
For example, if we fix a security level 95\% we can calculate the
VaR on that level, which is defined by
\begin{equation}
\label{eqns}
\text{VaR}_{0.95}\bigg(C_{i,J}
-E\left[C_{i,J}|\mathcal{D}_I\right]\bigg| \mathcal{D}_I\bigg) =
\min \left\{ x;~ P \bigg[
C_{i,J}-E\left[C_{i,J}|\mathcal{D}_I\right] > x \bigg|
\mathcal{D}_I \bigg] \le 0.05 \right\}.
\end{equation}

\section{Bootstrap and mean square error of prediction}
\label{section:Bootstrap DFCL}

Assume that we have calculated the Bayesian predictor or the CL
predictor given in \eqref{BayesPredictor}. Then we would like to
determine the prediction uncertainty, that is, we would like to study
the deviation of $C_{i,J}$ around its predictor. If one is only
interested in second moments, the so-called conditional mean
square error of prediction (MSEP), one can often estimate the
error terms analytically. However, other uncertainty measures like
Value-at-Risk (VaR) can only be determined numerically; see
(\ref{eqns}).

A popular numerical method is the bootstrap method. The bootstrap
technique was developed by Efron \cite{Efron} and extended by
Efron-Tibshirani \cite{ET} and Davison-Hinkley \cite{Davison}. In the actuarial literature the development of bootstrap procedures includes the work of Taylor \cite{Taylor87}, Taylor-McGuire \cite{Taylor05}, \cite{Taylor07}, England-Verrall \cite{EV99}, \cite{EV07} and Pinheiro et al. \cite{Pin03}.

This procedure allows one to obtain information regarding an
aggregated distribution given a single realisation of the data. To
apply the bootstrap procedure one introduces a minimal amount of
model structure such that resampling observations can be achieved
using observed samples of the data.

In this section we present a bootstrap algorithm in the classical
frequentist approach. That is, we assume that the CL factors
$\bm{F}=\bm{f}$ and the standard deviation parameters
$\bm{\Xi}=\bm{\sigma}$ given in Model Assumptions \ref{model ass}
are unknown constants. The bootstrap then generates synthetic data
denoted by $\mathcal{D}_{I }^{\ast }$ that allow for the study of
the fluctuations  of $\widehat{\bm{f}}^{(CL)}$ and
$\widehat{\bm{\sigma}}^{2(CL)}$ (for details see Section 7.4 in W\"uthrich-Merz
\cite{Wuthrich}). In the presented text we restrict
ourselves to the conditional resampling approach presented in
Section 7.4.2 of W\"uthrich-Merz \cite{Wuthrich}.

\noindent
\hrulefill%

\subsection{Non-parametric classical bootstrap (conditional
version)}

\begin{enumerate}
\label{page bootstrap}
\item Calculate estimated residuals $\widetilde{\varepsilon }_{i,j}$ for $i+j\leq I$, $j>0$, conditional on the estimators $\widehat{f}_{0:J-1}^{(CL)}$ and
$\widehat{\sigma}_{0:J-1}^{2(CL)}$ and the observed data
$\mathcal{D}_{I}$:
\begin{equation*}
\widetilde{\varepsilon }_{i,j }
~=~\widetilde{\varepsilon }_{i,j }(\widehat{f}%
_{j-1}^{(CL) },\widehat{\sigma }_{j-1}^{(CL)})
~=~\frac{C_{i,j}-\widehat{f}%
_{j-1}^{(CL) }C_{i,j-1}}{\widehat{\sigma }_{j-1}^{(CL)
}\sqrt{C_{i,j-1}}}.
\end{equation*}

\item These residuals $(\widetilde{\varepsilon}_{i,j})_{i+j\le I}$ give the
empirical bootstrap distribution
$\widehat{F}_{\mathcal{D}_{I}}$.%

\item Sample i.i.d. residuals $\widetilde{\varepsilon}_{i,j
}^{\ast }\sim \widehat{F}_{\mathcal{D}%
_{I}}$ for $i+j\leq I$, $j>0$.

\item Generate bootstrap observations (conditional resampling)
\begin{equation*}
C_{i,j}^{\ast }=\widehat{f}_{j-1}^{(CL)}C_{i,j-1}+\widehat{\sigma }_{j-1}^{(CL)}\sqrt{C_{i,j-1}}%
\widetilde{\varepsilon}_{i,j}^{\ast },
\end{equation*}%
which defines $\mathcal{D}_I^\ast~=~\mathcal{D}_I^\ast
(\widehat{\bm{f}}^{(CL) },\widehat{\bm{\sigma}}^{(CL)}) $. Note
that for the unconditional version of bootstrap we should generate
$C_{i,j}^{\ast}=\widehat{f}_{j-1}^{(CL)}C_{i,j-1}^{\ast}+\widehat{\sigma }_{j-1}^{(CL)}\sqrt{C_{i,j-1}^{\ast}}%
\widetilde{\varepsilon}_{i,j}^{\ast }$. For a discussion on this
approach, see Section 7.4.1 of \cite{Wuthrich}.

\item Calculate bootstrapped CL parameters
$\widehat{f}^{\ast}_{j}$and $\widehat{\sigma }_{j}^{2\ast }$ by
\begin{eqnarray*}
\widehat{f}^{\ast}_{j}&=&\frac{\sum\nolimits_{i=0}^{I-j-1}C_{i,j+1}^\ast}{%
\sum\nolimits_{i=0}^{I-j-1}C_{i,j}},\\
\widehat{\sigma }_{j}^{2\ast } &=&
\frac{1}{I-j-1}\sum\nolimits_{i=0}^{I-j-1}C_{i,j}\left( \frac{C^\ast_{i,j+1}}{%
C_{i,j}}-\widehat{f}_{j}^{\ast}\right) ^{2}.
\end{eqnarray*}
\item Repeat steps 3-5 and obtain empirical distributions from
the bootstrap samples $\widehat{C}_{i,J}^\ast$,
$\widehat{f}^{\ast}_{j}$ and $\widehat{\sigma }_{j}^{2\ast }$.
These are then used to quantify the parameter estimation
uncertainty.
\end{enumerate}

\noindent
\hrulefill%

This non-parametric classical bootstrap method can be seen as a
frequentist approach. This means that we do not express our
parameter uncertainty by the choice of an appropriate prior
distribution. We rather use a point estimator for the unknown
parameters and then study the possible fluctuations of this point
estimator.

The main difficulty now is that the non-parametric bootstrap
method, as described above, underestimates the ``true''
uncertainty. This comes from the fact that the estimated residuals
$\widetilde{\varepsilon}_{i,j}$, in general, have variance smaller
than 1 (see formula (7.23) in W\"uthrich-Merz \cite{Wuthrich}).
This means that our estimated residuals are not appropriately
scaled. Therefore, frequentists use several different scalings to
correct this fact (see formula (7.24) in W\"uthrich-Merz
\cite{Wuthrich} or England-Verrall \cite{EV}). Here, we use a
different approach by introducing the novel Bayesian bootstrap
method embedded within an MCMC-ABC algorithm to obtain empirically
the posterior distribution of the Bayesian DFCL model, described
below. Having obtained this, we can then calculate all required
Bayesian parameter estimates, capital reserve estimates and
associated risk measures such as VaR. Before presenting the
methodology for this novel MCMC-ABC algorithm we will finalize
this section with the decompositions of the MSEP under
frequentist, Bayesian and credibility approaches.

\subsection{Frequentist bootstrap estimates}

Let us for the time-being concentrate on the conditional MSEP
given by
\begin{eqnarray}
\text{msep}_{C_{i,J}|\mathcal{D}_I}\left(\widehat{C}_{i,J}\right)
&=&E\left[\left.\left(C_{i,J}-\widehat{C}_{i,J}\right)^2 \right|
\mathcal{D}_I\right]\\
\nonumber &=&\text{Var}\left(\left.C_{i,J}\right|
\mathcal{D}_I\right) +\left(E\left[\left.C_{i,J} \right|
\mathcal{D}_I\right]-\widehat{C}_{i,J}\right)^2.
\end{eqnarray}
The first term is known as the conditional process variance and
the second term as the parameter estimation uncertainty. In the
frequentist approach (i.e.~for given deterministic
$\bm{F}=\bm{f}$ and $\bm{\Xi}=\bm{\sigma}$) these terms can be
calculated as 
\begin{equation}\label{process variance}
\text{Var}\left(\left.C_{i,J}\right| \mathcal{D}_I\right)= \bigl(
E\left.\left[C_{i,J}\right|C_{i,I-i}\right] \bigr)^2~
\sum_{j=I-i}^{J-1}~
\frac{\sigma^2_j/f^2_{j}}{E\left.\left[C_{i,j}\right|C_{i,I-i}\right]}
\stackrel{def.}{=}C_{i,I-i} \Gamma_{I-i},
\end{equation}
and
\begin{equation}\label{parameter estimation error}
\left(E\left[\left.C_{i,J} \right|
\mathcal{D}_I\right]-\widehat{C}_{i,J}\right)^2
=C^2_{i,I-i}\left(\prod_{j={I-i}}^{J-1}f_j
-\prod_{j={I-i}}^{J-1}\widehat{f}^{(CL)}_j\right)^2
\stackrel{def.}{=}C_{i,I-i}^2 \Delta_{I-i};
\end{equation}
see W\"uthrich-Merz \cite{Wuthrich}, Section 3.2.

The process variance \eqref{process variance} is estimated by
replacing the parameters by its estimators,
\begin{equation}\label{process variance estimate}
\widehat{\text{Var}}\left(\left.C_{i,J}\right|
\mathcal{D}_I\right)= \bigl( \widehat{C}_{i,J} \bigr)^2~
\sum_{j=I-i}^{J-1}~
\frac{\widehat{\sigma}^{2(CL)}_j/(\widehat{f}^{(CL)}_j)^2}{\widehat{C}_{i,j}}
\stackrel{def.}{=}C_{i,I-i} \widehat{\Gamma}_{I-i}^{freq}.
\end{equation}
The parameter estimation error is more involved and there we need
the bootstrap algorithm. Assume that the bootstrap method gives
$T$ bootstrap samples $\widehat{f}_{j}^{\ast(1)}, \ldots,
\widehat{f}_{j}^{\ast(T)}$. Then the parameter estimation error
\eqref{parameter estimation error} is estimated by the sample
variance of the product of the bootstrap observation chain ladder
parameter estimates $\widehat{f}_{j}^{\ast(1)}, \ldots,
\widehat{f}_{j}^{\ast(T)}$, which gives the estimator $C_{i,I-i}^2
\widehat{\Delta}_{I-i}^{freq}$.

\subsection{Bayesian estimates}

In the Bayesian setup, (i.e.~choosing prior distributions for the
unknown parameters $\bm{F}$ and $\bm{\Xi}$) we obtain a natural
decomposition of the conditional MSEP:
\begin{eqnarray}
\text{msep}_{C_{i,J}|\mathcal{D}_I}\left(E\left[\left.C_{i,J}
\right| \mathcal{D}_I\right]\right)
&=&\text{Var}\left(\left.C_{i,J}\right|
\mathcal{D}_I\right)\\
\nonumber &=&E\left[\left.\text{Var}\left(\left.C_{i,J} \right|
\mathcal{D}_I, \bm{F},\bm{\Xi}\right)\right| \mathcal{D}_I\right]
+ \text{Var}\left(\left.E\left[\left.C_{i,J} \right|
\mathcal{D}_I,\bm{F},\bm{\Xi}\right]\right| \mathcal{D}_I\right).
\end{eqnarray}
The average process variance is given by (see W\"uthrich-Merz
\cite{Wuthrich}, Lemma 3.6)
\begin{eqnarray}\label{msepterm1}
&& E\left[\left.\text{Var}\left(\left.C_{i,J} \right|
\mathcal{D}_I, \bm{F},\bm{\Xi}\right)\right| \mathcal{D}_I\right]
=C_{i,I-i} \sum_{j=I-i}^{J-1} E\left[\left. \prod_{m=I-i}^{j-1}F_m
~\Xi_j^2 \prod_{n=j+1}^{J-1}
F_n^2 \right|\mathcal{D}_I\right]\\
&&\quad=C_{i,I-i} \sum_{j=I-i}^{J-1}
 \prod_{m=I-i}^{j-1}E\left[\left.F_m  \right|\mathcal{D}_I\right]
E\left[\left.\Xi_j^2 \right|\mathcal{D}_I\right]
\prod_{n=j+1}^{J-1} E\left[\left.F_n^2
\right|\mathcal{D}_I\right]\stackrel{def.}{=}C_{i,I-i}
\widehat{\Gamma}_{I-i}^{Bayes}, \nonumber
\end{eqnarray}
where we have used posterior independence
(\ref{PosteriorIndependence}). The parameter estimation error is
given by
\begin{equation}
\text{ Var}\left(\left.E\left[\left.C_{i,J} \right|
\mathcal{D}_I,\bm{F},\bm{\Xi}\right]\right| \mathcal{D}_I\right)
=C_{i,I-i}^2 \text{ Var}\left( \left.\prod_{j=I-i}^{J-1} F_j
\right| \mathcal{D}_I\right)\stackrel{def.}{=}C_{i,I-i}^2
\widehat{\Delta}_{I-i}^{Bayes},
\end{equation}
where we have used \eqref{BayesPredictor}. Using
(\ref{PosteriorIndependence}), we obtain for the last term
\begin{equation}\label{msepterm2}
C_{i,I-i}^2 \widehat{\Delta}_{I-i}^{Bayes} =C_{i,I-i}^2 \left[
\prod_{j=I-i}^{J-1}E\left[ \left. F^2_j \right|
\mathcal{D}_I\right] -\prod_{j=I-i}^{J-1}E\left[ \left. F_j
\right| \mathcal{D}_I\right]^2 \right].
\end{equation}
In order to calculate these two terms given in \eqref{msepterm1}
and \eqref{msepterm2}, we need to calculate the posterior
distribution of $(\bm{F},\bm{\Xi})$, given $\mathcal{D}_I$. Since
we do not have a full distributional model, we cannot write down
the likelihood function, which would allow for analytical
solutions or Markov chain Monte Carlo (MCMC) simulations.
Therefore we introduce the ABC framework which allows for
distribution-free simulations using appropriate bootstrap samples
and a distance metric. This will be discussed in
Section \ref{ABCSection}.

\subsection{Credibility Estimates}
As mentioned previously, we can also consider the credibility
estimates given in Gisler-W\"uthrich \cite{GW}. As long as we are
only interested in the second moments (i.e.~conditional MSEP) we
can also use credibility estimators, which are minimum variance
estimators that are linear in the observations. For diffuse priors
we obtain the approximation given in Corollary 7.2 of
Gisler-W\"uthrich \cite{GW}
\begin{equation}
\widehat{\text{msep}}_{C_{i,J}|\mathcal{D}_I}\left(E\left[\left.C_{i,J}
\right|
\mathcal{D}_I\right]\right) =C_{i,I-i}\widehat{\Gamma }%
_{I-i}^{cred}+C_{i,I-i}^{2}\widehat{\Delta }_{I-i}^{cred},
\end{equation}%
where%
\begin{eqnarray}
\widehat{\Gamma }_{I-i}^{cred} &=&\sum_{j=I-i}^{J-1}
\left\{ \prod_{m=I-i}^{j-1}%
\widehat{f}_{m}^{(CL)}~ \widehat{\sigma }_{j}^{2(CL)}~
\prod_{n=j+1}^{J-1}\left( (\widehat{f}_{n}^{(CL)})^{2}+
\frac{\widehat{\sigma }_{n}^{2(CL)}}{\sum_{i=0}^{I-n-1}C_{i,n}}
\right) \right\} ,  \label{Hilf21} \\
\widehat{\Delta }_{I-i}^{cred}
&=&\prod_{j=I-i}^{J-1}\left((\widehat{f}_{j}^{(CL)})^{2}+%
\frac{\widehat{\sigma
}_{j}^{2(CL)}}{\sum_{i=0}^{I-j-1}C_{i,j}}\right)
-\prod_{j=I-i}^{J-1}(\widehat{f}_{j}^{(CL)})^{2}.  \label{Hilf22}
\end{eqnarray}%
In the results section we compare the frequentist bootstrap
approach, the credibility approach and the ABC bootstrap approach
that is described below (see Table 7 below).




\section{ABC for intractable likelihoods and numerical Markov chain sampler}
\label{ABCSection}
To estimate numerically the parameters, predicted claims and
associated uncertainty measures such as the MSEP presented in the
previous sections, the Bayesian approach requires the ability to
sample from the posterior distribution of the DFCL model
parameters. Obtaining samples $\left\{ \bm{f}^{\left( t\right)
},\bm{\sigma}^{2\left( t\right) }\right\} _{t=1:T}$ which are
realisations of a random vector distributed with a posterior
distribution $\pi \left(\bm{f},\bm{\sigma} |\mathcal{D}_{I}\right)
$ in the DFCL model is difficult since the likelihood is
intractable. Hence, standard numerical approaches such as Markov
chain Monte Carlo (MCMC) algorithms (see Gilks et
al.~\cite{Gilks}) cannot be directly used since they all require
explicit repeated evaluation of the likelihood function at each
stage of the Markov chain sampling algorithm. It is common to
avoid this difficulty by making distributional assumptions for the
form of the likelihood. This then violates the DFCL model
assumption but allows for relatively standard sampling procedures
to be applied. In this regard, one possible approach involves
making a specific Gaussian assumption for the likelihood. One
problem with this assumption, which is evident immediately, is that
it precludes skewness in the model. Here, we do not make any such
assumptions and instead we work in a truly distribution-free model
using ABC to facilitate sampling from an intractable posterior
distribution.

There is an additional complexity in the DFCL model not typically
encountered when working with ABC methodology. Typically, ABC
methodology is developed in the case in which the model likelihood
cannot be evaluated point-wise, but conditional on parameter
values, synthetic data is easily simulated from the model; see
examples in Peters-Sisson~\cite{Peters1a} and Peters et
al.~\cite{Peters1f}. This is not the case in the Bayesian DFCL
model. Under the DFCL model the likelihood is only expressed by
moment conditions, hence we cannot evaluate the likelihood
point-wise and also the simulation from the likelihood cannot be
performed directly. This is why we introduce the novel concept of
the Bayesian bootstrap which is embedded within the ABC
methodological framework.

Hence, to sample from the posterior in our DFCL model we develop a
novel formulation of the ABC methodology based on the bootstrap
and conditional back transformation procedure, similar to that
discussed in Section \ref{section:Bootstrap DFCL}.

ABC methods aim to sample from posterior distributions in the
presence of computationally intractable likelihood functions. For
an application in risk modelling of ABC methodology, see
Peters-Sisson \cite{Peters1a}. In this article we present a novel
MCMC-ABC algorithm. Before presenting some details of the
numerical MCMC procedure, we note that alternative numerical
algorithms could be considered in the ABC context. For example, a
sequential Monte Carlo (SMC) based algorithms which can improve
simulation efficiency can be found in Del Moral et
al.~\cite{DelMoral1a}, Sisson et al.~\cite{Sisson1a}, Peters et
al.~\cite{Peters1b},\cite{Peters1c} and Marjoram et
al.~\cite{Marjoram}.

\subsection{ABC methodology}
In this section we provide a brief description of ABC methodology,
which describes a suite of methods developed specifically for
working with models in which the likelihood is computationally
intractable. Here we work with a Bayesian model and consider the
likelihood intractability to arise in the sense that we may not
evaluate the likelihood point-wise.

The ABC method we consider here embeds an intractable target
posterior distribution, in our case denoted by $\pi
\left(\bm{f},\bm{\sigma} |\mathcal{D}_{I}\right)$, into a general
augmented model
\begin{align}
\pi
\left(\bm{f},\bm{\sigma},\mathcal{D}^{\ast}_{I},\mathcal{D}_{I}\right)
=\pi\left(\mathcal{D}_{I}|\mathcal{D}^{\ast}_{I},\bm{f},\bm{\sigma}\right)
\pi\left(\mathcal{D}^{\ast}_{I}|\bm{f},\bm{\sigma}\right)\pi\left(\bm{f},\bm{\sigma}\right),
\end{align}
where $\mathcal{D}^{\ast}_{I}$ is an auxiliary vector on the same
space as $\mathcal{D}_{I}$. In this augmented Bayesian model, the
weighting function
$\pi\left(\mathcal{D}_{I}|\mathcal{D}^{\ast}_{I},\bm{f},\bm{\sigma}\right)$
weights the intractable posterior. In this paper we consider the
hierarchical model assumption, where we work with
$\pi\left(\mathcal{D}_{I}|\mathcal{D}^{\ast}_{I},\bm{f},\bm{\sigma}\right)
= g\left(\mathcal{D}_{I}|\mathcal{D}^{\ast}_{I}\right)$; see
Reeves and Pettitt~\cite{Reeves}.

The mechanism in the ABC framework which allows one to avoid the
evaluation of the intractable likelihood involves replacing this
evaluation with data simulation from the likelihood. That is,
given a realisation of the parameters of the model, a synthetic
data set $\mathcal{D}^{\ast}_{I}$ is generated and compared to
the original data set. This is a key aspect of the novel
methodology we develop in this paper, since we utilise a bootstrap
procedure to perform this simulation in the DFCL model setting.

Then summary statistics $S(\mathcal{D}^{\ast}_{I})$ derived from
this data are compared to summary statistics of the observed data
$S(\mathcal{D}_{I})$ and a distance $\rho\left(S(\mathcal{D}^{\ast}_{I}),S(\mathcal{D}_{I})\right)$ is calculated.
Finally, a weight is given to these parameters according to the
weighting function
$g\left(\mathcal{D}_{I}|\mathcal{D}^{\ast}_{I}\right)$, which may
give greater weight when $S(\mathcal{D}^{\ast}_{I})$ and
$S(\mathcal{D}_{I})$ are close (i.e. where
$\rho\left(S(\mathcal{D}^{\ast}_{I}),S(\mathcal{D}_{I})\right)$ is
small).

For example, under the ``Hard Decision'' (HD) weighting given by
\begin{align}
\label{WeightingFunctionHD}
g\left(\mathcal{D}_{I}|\mathcal{D}^{\ast}_{I}\right) &\propto
\begin{cases}
{1} & \text{if }\rho\left(S\left(\mathcal{D}_{I}\right),S\left(\mathcal{D}^{\ast}_{I}\right)\right)\leq\epsilon, \\
{0} & \text{ otherwise;}%
\end{cases}
\end{align}
\noindent a reward is given to summary statistics of the augmented
auxiliary variables $S\left(\mathcal{D}^{\ast}_{I}\right)$
within an $\epsilon$-tolerance of the summary statistic of the
actual observed data $S\left(\mathcal{D}_{I}\right)$, as measured
by distance metric $\rho$.

Hence, in the ABC context, an approximation to the intractable
target posterior marginal distribution $\pi
\left(\bm{f},\bm{\sigma} |\mathcal{D}_{I}\right)$, for which we
are interested in formulating an empirical estimate, is given by
\begin{align}
\begin{split}
\label{postABC} \pi_{ABC}\left(\bm{f},\bm{\sigma}
|\mathcal{D}_{I},\epsilon\right) & \propto \int
g\left(\mathcal{D}_{I}|\mathcal{D}^{\ast}_{I}\right)\pi\left(\mathcal{D}^{\ast}_{I}|\bm{f},\bm{\sigma}
\right)\pi\left(\bm{f},\bm{\sigma}\right) d\mathcal{D}^{\ast}_{I}.
\end{split}
\end{align}
As briefly mentioned, obtaining samples from the ABC posterior can
be achieved using a number of numerical procedures, in this paper
we consider an MCMC approach. The MCMC class of likelihood-free
algorithm is justified on a joint space formulation, in which the
stationary distribution of the Markov chain is given by
$\pi_{ABC}\left(\bm{f},\bm{\sigma},\mathcal{D}^{\ast}_{I}|\mathcal{D}_{I},\epsilon\right)
$. The corresponding target distribution for the marginal
distribution $\pi_{ABC}\left(\bm{f},\bm{\sigma}
|\mathcal{D}_{I},\epsilon\right)$ is then obtained via numerical
integration. Note that the marginal posterior distribution
$\pi_{ABC}\left(\bm{f},\bm{\sigma}
|\mathcal{D}_{I},\epsilon\right) \rightarrow
\pi\left(\bm{f},\bm{\sigma} |\mathcal{D}_{I}\right)$ as $\epsilon
\rightarrow 0$, recovering the "true" (intractable) posterior,
assuming that $S\left(\mathcal{D}_{I}\right)$ are sufficient
statistics and that the weighting function converges to a point
mass on $S\left(\mathcal{D}_{I}\right)$ as $\epsilon \rightarrow
0$; see Peters-Sisson~\cite{Peters1a} and references therein
for detailed discussion. Accordingly, the tolerance $\epsilon$ is
typically set as low as possible for a given computational budget.
In this paper we focus on the class of MCMC-based sampling
algorithms.

The ABC methodology is novel both in the statistics literature and
in the actuarial literature. It is informative to clearly
provide the justification for this approach both theoretically and
numerically. The simplest understanding of ABC is achieved by
considering a rejection algorithm, therefore we provide a basic
argument for how the ABC methodology works in simple rejection
sampling in Appendix \ref{section: Justification for ABC}. The
actuarial DFCL model considered in this paper requires the more
sophisticated MCMC-ABC methodology described below.

\subsection{Technical justification for MCMC-ABC algorithm}

For given observations $\mathcal{D}_{I}$ we want to sample from
$\pi_{ABC}(\bm{f},\bm{\sigma}|\mathcal{D}_{I})$ with an
intractable likelihood function. We assume that
$S(\mathcal{D}_{I})$ is either the data itself or a summary of the
data such as a sufficient statistic for the model from which we
assume data $\mathcal{D}_{I}$ is a realisation. We assume that,
given a set of parameters values $\left(\bm{f},\bm{\sigma}\right)$,
we can generate from the DFCL model (via a conditional bootstrap
procedure) a synthetic data set denoted $\mathcal{D}_{I}^{\ast}$.
We define a hard decision function
$g(\mathcal{D}_{I}^{\ast},\mathcal{D}_{I})=\mathbb{I}\{\rho(S(\mathcal{D}_{I}^{\ast}),S(\mathcal{D}_{I}))<\epsilon\}(\mathcal{D}_{I}^{\ast})$
for a given tolerance level $\epsilon>0$ and a distance metric
$\rho (\cdot,\cdot)$, where $\mathbb{I}\{\cdot\}$ is the indicator
function which equals 1 if the event is true and 0 otherwise. As
demonstrated in Appendix A, we use the approximation,
\eqref{approx2}-\eqref{approx3}, which gives us in the Bayesian
DFCL model setting,
\begin{equation}
\label{approx4est}
\pi_{ABC}(\bm{f},\bm{\sigma}|\mathcal{D}_{I},\epsilon) =
\frac{\int g(\mathcal{D}_{I}|\mathcal{D}_{I}^{\ast})
~\pi(\mathcal{D}_{I}^{\ast}|\bm{f},\bm{\sigma})~\pi(\bm{f},\bm{\sigma})~d\mathcal{D}_{I}^{\ast}}{\int
g(\mathcal{D}_{I}|\mathcal{D}_{I}^{\ast})
~\pi(\mathcal{D}_{I}^{\ast}|\bm{f},\bm{\sigma})~\pi(\bm{f},\bm{\sigma})~d\mathcal{D}_{I}^{\ast}~d\bm{f}~d\bm{\sigma}}
=\frac{\pi(\bm{f},\bm{\sigma})E\left.\left[g(\mathcal{D}_{I}|\mathcal{D}_{I}^{\ast})\right|\bm{f},\bm{\sigma}\right]}
{E\left[g(\mathcal{D}_{I}|\mathcal{D}_{I}^{\ast})\right]}.
\end{equation}
In the next step the numerator of \eqref{approx4est} is
approximated using the empirical distribution:
\begin{equation}
\pi(\bm{f},\bm{\sigma})E\left.\left[g(\mathcal{D}_{I}|\mathcal{D}_{I}^{\ast})\right|\bm{f},\bm{\sigma}\right]
\approx \pi(\bm{f},\bm{\sigma}) \frac{1}{L} \sum_{l=1}^L g\left(
\mathcal{D}_{I}|\mathcal{D}_{I}^{\ast,(l)}(\bm{f},\bm{\sigma})\right),
\end{equation}
where
$\mathcal{D}_{I}^{\ast,(l)}(\bm{f},\bm{\sigma})\stackrel{i.i.d.}{\sim}
\pi(\mathcal{D}_{I}^{\ast}|\bm{f},\bm{\sigma})$. Finally, we need
to consider the denominator $E\left[g(X|y)\right]$. In general
this has a non-trivial form that cannot be calculated
analytically. However, since we use an MCMC based method the
denominators cancel in the accept-reject stage of the algorithm.
Therefore, the intractability of the denominator does not impede
sampling from the posterior. Thus we use
\begin{equation}\label{approx4}
\begin{aligned}
\pi_{ABC}(\bm{f},\bm{\sigma}|\mathcal{D}_{I},\epsilon) &\approx
\frac{\int g(\mathcal{D}_{I}|\mathcal{D}_{I}^{\ast})
~\pi(\mathcal{D}_{I}^{\ast}|\bm{f},\bm{\sigma})~\pi(\bm{f},\bm{\sigma})~d\mathcal{D}_{I}^{\ast}}{\int
g(\mathcal{D}_{I}|\mathcal{D}_{I}^{\ast})~\pi(\mathcal{D}_{I}^{\ast}|\bm{f},\bm{\sigma})~\pi(\bm{f},\bm{\sigma})~d\mathcal{D}_{I}^{\ast}~d\bm{f}~d\bm{\sigma}}\\
&\propto\pi(\bm{f},\bm{\sigma})E\left.\left[g(\mathcal{D}_{I}|\mathcal{D}_{I}^{\ast})\right|\bm{f},\bm{\sigma}\right]\\
&\approx \pi(\bm{f}) \pi(\bm{\sigma}) \frac{1}{L} \sum_{l=1}^L
g\left(\mathcal{D}_{I}|\mathcal{D}_{I}^{\ast,(l)}(\bm{f},\bm{\sigma})\right)
\end{aligned}
\end{equation}
in order to obtain samples from
$\pi_{ABC}(\bm{f},\bm{\sigma}|\mathcal{D}_{I},\epsilon)$. Almost
universally, $L=1$ is adopted to reduce computation but on the
other hand this will slow down the rate of convergence to the
stationary distribution.

Note that sometimes one also uses softer decision functions for
$g(\cdot|\cdot)$. The role of the distance measure $\rho $ is
evaluated by Peters et al.~\cite{Peters1f}. We further extend this
analysis to the class of models considered in this paper. We
analyse several choices for the distance measure $\rho $ such as
Mahlanobis distance, scaled Euclidean distance and the Manhattan
``City Block'' distance. Fan et al.~\cite{Fan} demonstrate that it
is not efficient to utilise the standard Euclidean distance,
especially when summary statistics considered are on different
scales.

Additionally, using an MCMC-ABC algorithm, it is important to
assess convergence diagnostics. Particularly when using MCMC-ABC
where serial correlation in the Markov chain samples can be
significant if the sampler is not designed carefully. We assess
autocorrelation of the simulated Markov chain, the Geweke
\cite{Geweke1991} time series statistic and the Gelman-Rubin
\cite{Gelman1a} R-statistic convergence diagnostic in an ABC
setting.

{\bf Concluding:} We apply three different techniques in order to
treat the intractable likelihood:
\begin{enumerate}
\item{ABC is used to get a handle on the likelihood and therefore the
intractable posterior.}
\item{As a result of using ABC we need to be able to
generate synthetic data samples from the DFCL model given
realisations of the parameters. These data samples come from the bootstrap
algorithm.}
\item{We use a well understood MCMC based sampling
algorithm that does not require calculation of the non-analytic
normalizing constants for the target distribution
$\pi_{ABC}(\bm{f},\bm{\sigma}|\mathcal{D}_{I},\epsilon)$. The
reason for this is that in the acceptance probability of the MCMC
algorithm, the normalizing constant for the target posterior
appears both in the numerator and denominator, resulting in
cancellation.}
\end{enumerate}
The specific details of the MCMC algorithm and ABC choices are
provided in the Appendix B.

\section{Example 1: Analysis of MCMC-ABC bootstrap methodology on synthetic data}
To test the accuracy of the methodology, first we use synthetic
data generated with known parameter values. The tuning of the
proposal distribution in this study is done for the simplest
``base'' distance metric, the weighted Euclidean distance. To study
the effect of the distance metric in a comparative fashion we
shall keep the proposal distribution unchanged.

The first example we present has a claims triangle of size
$I=J=9$. In this example we fix the true model parameters, denoted by $\bm{f}%
\bm{=}\left( f_{0},\ldots,f_{J-1}\right) $ and $\bm{\sigma }^{2}%
\bm{=}\left( \sigma _{0}^{2},\ldots,\sigma _{J-1}^{2}\right) $ and
given in Table \ref{tab2}, used to generate the synthetic data
set.

\subsection{Generation of synthetic data}
To generate the synthetic observations for $\mathcal{D}_{I}$, we
generate randomly the first column (i.e. $\mathcal{B}_{0}$).
Then conditional on this realisation of $\mathcal{B}_{0}$ we make
use of the model given in (\ref{model ass}) to generate the
remaining columns of $\mathcal{D}_{I}$, ensuring the model
assumptions are satisfied. This requires setting $C_{i,0}$
sufficiently large (for appropriate choices of $\bm{f}$ and
$\bm{\sigma }^{2}$) and then sampling i.i.d. realisations of
$\varepsilon_{i,j} \sim \mathcal{U}\left[-\sqrt{3},
\sqrt{3}\right]$ used to obtain $\mathcal{D}_I$; see the
observations in Table \ref{tab2}.

\subsection{Sensitivity analysis and convergence assessment}

We perform a sensitivity analysis, studying the impact of the
distance metric on the mixing of the Markov chain in the case of
joint estimation of the chain ladder factors and the variance
parameters.

The pre-tuned coefficient of variation of the Gamma proposal
distribution for each parameter of the posterior was performed
using the following settings; $T_{b}=50,000$, $\tilde{T}=200,000$,
$\epsilon^{\min}=0.1$ and initial values $\gamma_j = 1$ for all $j
\in \{1,\ldots,2J\}$. Additionally, the prior parameters for the
chain ladder factors $F_j$ were set as $\left(\alpha,\beta\right)
= \left(2,1.2/2\right)$ and the parameters for the variance
parameters $\Xi^{-2}_j$ were set as
$\left(a,b\right)=\left(2,1/2\right)$.

After tuning the proposal distributions during burn-in and
rounding the shape parameters, we found that $\gamma_j = 10$ for
all $j \in \{1,\ldots,2J\}$ produced average acceptance
probabilities for each parameter between 0.3 and 0.5. This is a
range typically used in practice when designing MCMC
sampling algorithms.

Then, keeping the proposal distribution constant and using a common
data set $\mathcal{D}_{I}$, we ran three versions of the MCMC-ABC
algorithm for 200,000 samples corresponding to:
\begin{enumerate}
\item{scaled Euclidean distance and joint estimation of posterior for
$\bm{F},\bm{\Xi}^2$;}
\item{Mahlanobis distance (modified) and joint estimation of posterior for
$\bm{F},\bm{\Xi}^2$; and}
\item{Manhattan ``City Block'' distance and joint estimation of posterior for
$\bm{F},\bm{\Xi}^2$.}
\end{enumerate}

\subsection{Convergence diagnostics}
We estimate the three convergence diagnostics given in
Appendix B. The results of this analysis are presented as a
function of Markov chain iteration $t$ post burn-in of 50,000
samples.

\begin{figure}[tbp]
\centerline{\includegraphics[width=.3\textheight]{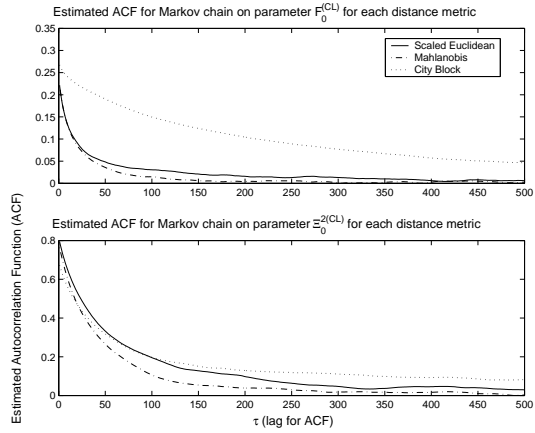}}
\caption{{\small{ Estimated Autocorrelation Function (ACF) for
parameters $F_0$ and $\Xi^2_0$}.}} \label{fig1}
\end{figure}

\textbf{Autocorrelation Function}: Figure \ref{fig1} shows the estimated autocorrelation functions for the Markov chains of the random variables $F_0$ and $\Xi_0^{2}$. We analyze
the marginal parameters to get a reasonable estimate of the mixing
behavior of the MCMC-ABC algorithm. The results demonstrate the
degree of serial correlation in the Markov chains generated for
these parameters as a function of lag time $\tau$. The higher the
decay rate in the tail of the estimated ACF as a function of
$\tau$, the better the mixing of the MCMC algorithm. Due to the
independence properties of this model there is little difference
between results obtained for Scaled Euclidean and Mahlanobis
distances. As shown in Appendix C, the estimate of the covariance
matrix is diagonal on all but the right lower $2 \times 2$ block.
Hence, we recommend using the simple Scaled Euclidean distance
metric as it provided the best trade-off between simplicity and
mixing performance.

\begin{figure}[tbp]
\centerline{\includegraphics[width=.3\textheight]{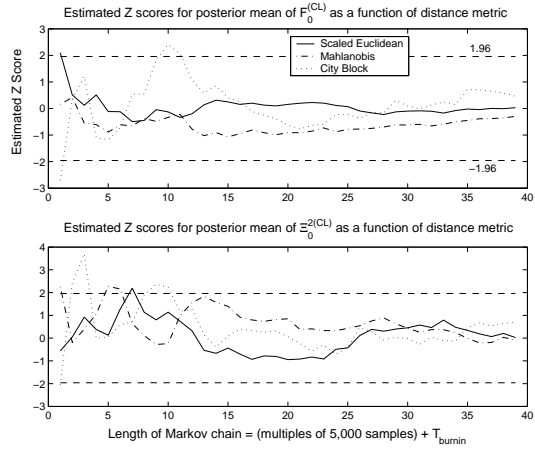}}
\caption{{\small{ Estimated Z scores for the posterior mean of
parameters $F_0$ and $\Xi^2_0$ as a function of the length of the
Markov chain $\widetilde{T}$.}}} \label{fig2}
\end{figure}

\textbf{Geweke Time Series Diagnostic}: Figure \ref{fig2} shows
results for the Geweke time series diagnostic. Again, we present
the results for the random variables $F_0$ and $\Xi_0^{2}$. Note,
we used the posterior mean as the sample function and a set of
increasing values for $\widetilde{T}$ from $T_{b}+5,000$ increasing in
steps of 5,000 samples to $T$. In each case we split the chain in
each ``window'' given by $\{\theta_i^{(t)}\}_{t=1:T_1}$ and
$\{\theta_i^{(t)}\}_{t=T^*:\widetilde{T}}$ according to
recommendations from Geweke et al. \cite{Geweke1991}. We then
calculate the convergence diagnostic $Z_{\widetilde{T}}$ which is the
difference between these two means divided by the asymptotic
standard error of their difference. As the chain length increases
$\widetilde{T} \rightarrow \infty$, the sampling distribution of
$Z \rightarrow {\cal N}(0,1)$ if the chain has converged. Hence
values of $Z_{\widetilde{T}}$ in the tails of a standard normal
distribution suggest that the chain was not fully converged early
on (i.e. during the 1st window). Hence, we plot $Z_{\widetilde{T}}$
scores versus increasing $\widetilde{T}$ and monitor if they lie
within a $95\%$ confidence interval $Z_{\widetilde{T}} \in [-1.96,
1.96]$. The results in Figure \ref{fig2} clearly demonstrate the convergence 
properties of the distance functions differ. Again
this is more material in the Markov chain for the variance
parameter when compared to the Markov chain results for the chain
ladder factor. The main point we note is that again one would
advise against use of the ``City block'' distance metric.

\begin{figure}[tbp]
\centerline{\includegraphics[width=.3\textheight]{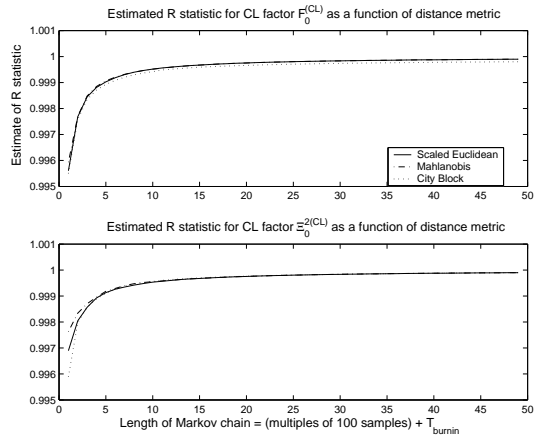}}
\caption{{\small{ Estimated R statistic for parameters $F_0$ and
$\Xi^2_0$ as a function of the length of the Markov chain
$\widetilde{T}$.}}} \label{fig3}
\end{figure}

\textbf{Gelman and Rubin R statistic}: Figure \ref{fig3}
presents the Gelman and Rubin convergence diagnostic. To calculate
this we ran 20 chains in parallel, each of length 10,000 samples
and for each chain we discarded 250 samples as burn-in. We then
estimated the $R$ statistic as a function of simulation time post
burn-in. Figure \ref{fig3} shows the convergence rate
of the $R$ statistic to 1 for each distance metric on increasing
blocks of 200 samples. Using this summary statistic, all three
distance metrics are very similar in terms of convergence rate of
the R statistic to 1.

Overall, these three convergence diagnostics demonstrate that the
simple scaled Euclidean distance metric is the superior choice.
Secondly, we see appropriate convergence of the Markov chains
under three convergence diagnostics which tests different aspects
of the mixing of the Markov chains, giving confidence in the
performance of the MCMC-ABC algorithm for this model.

\subsection{Bayesian parameter estimates}

In this section we present results for the scaled Euclidean
distance metric, with a Markov chain of length 200,000 samples
discarding the first 50,000 samples as burn-in. Table
\ref{tab3} shows the CL parameter estimates for the DFCL
model and the associated parameter estimation error. We define the
following quantities:
\begin{itemize}
\item{$\widehat{f}^{(MAP)}_{j}|\sigma_{0:J-1}$,
$\widehat{f}^{(MMSE)}_{j}|\sigma_{0:J-1}$,
$\widehat{\sigma}_{f_{j}}|\sigma_{0:J-1}$ and
$[\hat{q}_{0.05},\hat{q}_{0.95}]|\sigma_{0:J-1}$ denote
respectively the Maximum a-Posteriori, Minimum Mean Square Error,
posterior standard deviation of the conditional distribution of
chain ladder factor $F_j$ and the posterior coverage probability
estimates at $5\%$ of the conditional distribution of chain ladder
factor $F_j$. Each of these estimates is conditional on knowledge
of the true $\sigma_{0:J-1}$.}
\item{$\widehat{f}^{(MAP)}_{j}$, $\widehat{f}^{(MMSE)}_{j}$,
$\widehat{\sigma}_{f_{j}}$ and $[\hat{q}_{0.05},\hat{q}_{0.95}]$
denote the same quantities for the unconditional distribution
after joint estimation of $F_{0:J-1}$ and $\Xi_{0:J-1}$.}
\item{$Ave[A\left(\theta_{1:2J},f_{j}\right)]$ and $Ave[A\left(\theta_{1:2J},\sigma_{j}\right)]$  denote the average acceptance probabilities of the Markov chain.}
\item{$\widehat{\sigma}^{2(MAP)}_{j}$,
$\widehat{\sigma}^{2(MMSE)}_{j}$,
$\widehat{\sigma}_{\sigma^2_{j}}$ and
$[\hat{q}_{0.05},\hat{q}_{0.95}]$ denote the same quantities for
the chain ladder variances as those defined above for chain ladder
factors.}
\end{itemize}
Note, the estimates for $\widehat{f}^{(MAP)}_{j}$ and $\widehat{\sigma}^{(MAP)}_{j}$ were obtained marginally.
For the frequentist approach we obtain the
standard error in the estimates by using 1,000 bootstrap realisations of $\left\{ \mathcal{D}%
_{I}^{\left( s\right) }\right\} _{s=1:1,000}$ to obtain $\left\{
\widetilde{\bm{f}}_{\left( s\right) }^{\left( CCL\right)
},\widetilde{\bm{\sigma} }_{\left( s\right) }^{2\left( CCL\right)
}\right\} _{s=1:1,000}$. We use these bootstrap samples to
calculate the standard deviation in the estimates of the
parameters in the classical frequentist CL approach, given in
brackets $\left( .\right) $ next to their corresponding
estimators. The standard errors in the Bayesian parameter
estimates are obtained by blocking the Markov chain into 100
blocks of length 1,500 samples and estimating the posterior
quantities on each block.

\section{Example 2: Real Claims Reserving data}
In this example we consider estimation using real claims reserving
data from W\"uthrich-Merz \cite{Wuthrich}, see Table 3. This
yearly loss data is turned into annual cumulative claims and
divided by 10,000 for the analysis in this example. We use the
analysis from the previous study to justify use of the joint
MCMC-ABC simulation algorithm with a scaled Euclidean distance
metric.

We pre-tuned the coefficient of variation of the Gamma proposal
distribution for each parameter of the posterior. This was
performed using the following settings: $T_{b}=50,000$,
$\widetilde{T}=200,000$, $\epsilon^{\min}=10^{-5}$ and initial values
$\gamma_j = 1$ for all $j \in \{1, \ldots ,2J\}$. Here we make a
strict requirement of the tolerance level to ensure we have
accurate results from our ABC approximation. Additionally, the
prior parameters for the chain ladder factors $F_j$ were set as
$\left(\alpha_j,\beta_j\right) =
\left(1,\widehat{f}^{(CL)}_{j}\right)$ and the parameters for the
variance $\Xi^{-2}_j$ priors were set as
$\left(a_j,b_j\right)=\left(1,\widehat{\sigma}^{(CL)}_{j}\right)$.
The code for this problem was written in Matlab and it took
approximately 10 min to simulate 200,000 samples from the MCMC-ABC
algorithm on Intel Xeon 3.4GHz processor with 2Gb RAM.

After tuning the proposal distributions during burn-in we
obtained rounded shape parameters\\
$\gamma_{1:9} = [50; 100; 500; 500; 5,000; 20,000; 100,000;
2,000,000; 3,000,000]$ provided average acceptance probabilities
between 0.3 and 0.5.

\begin{figure}[tbp]
\centerline{\includegraphics[width=.4\textheight]{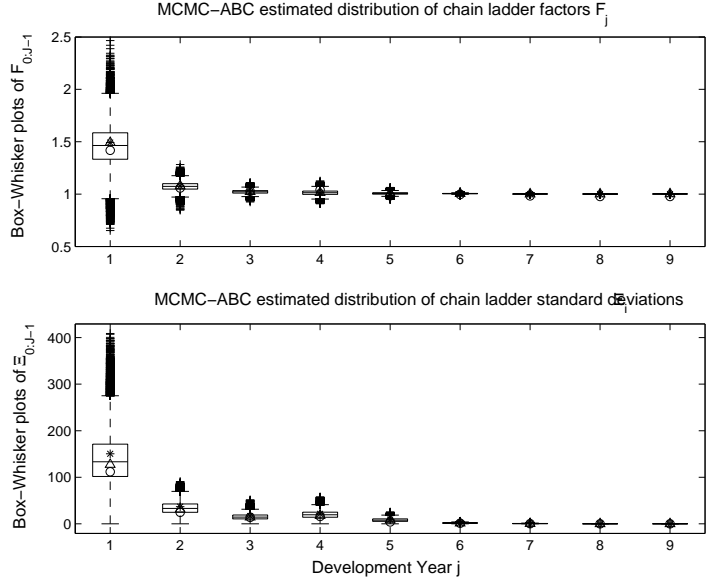}}
\caption{{\small{Box-Whisker plots of parameters $\bm{F}$ and
$\bm{\Xi}$ with each box marking the $25^{th},50^{th},75^{th}$
percentiles. Top: 200,000 MCMC-ABC samples to estimate posterior
for $\bm{F}$. The sample mean and mode are denoted by '*' and 'o'
respectively. The classical estimators $\widehat{\bm{f}}^{(CL)}$
are denoted by $\triangle$. Bottom: 200,000 MCMC-ABC samples to
estimate posterior for $\bm{\Xi}$. The sample mean and mode are
denoted by '*' and 'o' respectively. The classical estimators
$\widehat{\bm{\sigma}}^{(CL)}$ are denoted by '$\triangle$'.}}}
\label{fig4}
\end{figure}

\begin{figure}[tbp]
\centerline{\includegraphics[width=.4\textheight]{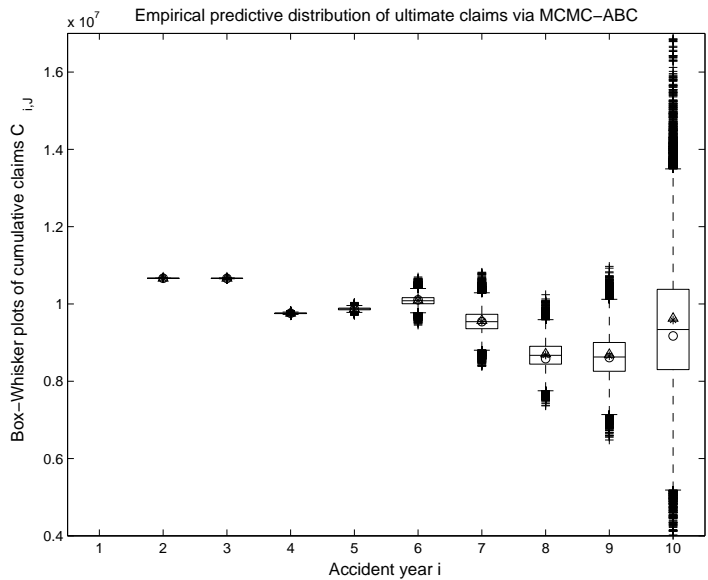}}
\caption{{\small{Box-Whisker plots of predictive distribution of
cumulative ultimate claims $C_{1:J}$ with the box marking the
$25^{th},50^{th},75^{th}$ percentiles; see also Table 6. The mean
predicted ultimate claims under a Bayesian approach (using MMSE
point estimates) are marked with '*', the predicted mode for the
ultimate claims (using MAP point estimates) is marked with 'o' and
the mean predicted ultimate claims under the DFCL classical method
are marked with '$\triangle$'.}}} \label{fig5}
\end{figure}

\textbf{Estimates of $\bm{f}$ and $\bm{\sigma}$} \\
Figures \ref{fig4} presents box-whisker plots of
estimates of the distributions of the parameters $F_{0:J-1}$ and $\Xi_{0:J-1}$ 
obtained from the MCMC-ABC algorithm, post burn-in.
Figure \ref{fig5} shows the Bayesian MCMC-ABC empirical
distributions of the ultimate claims, $C_{i,J}$ for
$i=1,\ldots,I$. In Table \ref{tab5} we present the predicted
cumulative claims for each year along with the estimates for the
chain ladder factors and chain ladder variances under both the
classical approach and the Bayesian model. We see that with this
fairly vague prior specified, we do indeed obtain convergence of
the MCMC-ABC based Bayesian estimates
$\widehat{\bm{f}}^{(MMSE)},\widehat{\bm{\sigma}}^{(MMSE)}$ to the
classical estimates
$\widehat{\bm{f}}^{(CL)},\widehat{\bm{\sigma}}^{(CL)}$.

\begin{figure}[tbp]
\centerline{\includegraphics[width=.6\textheight]{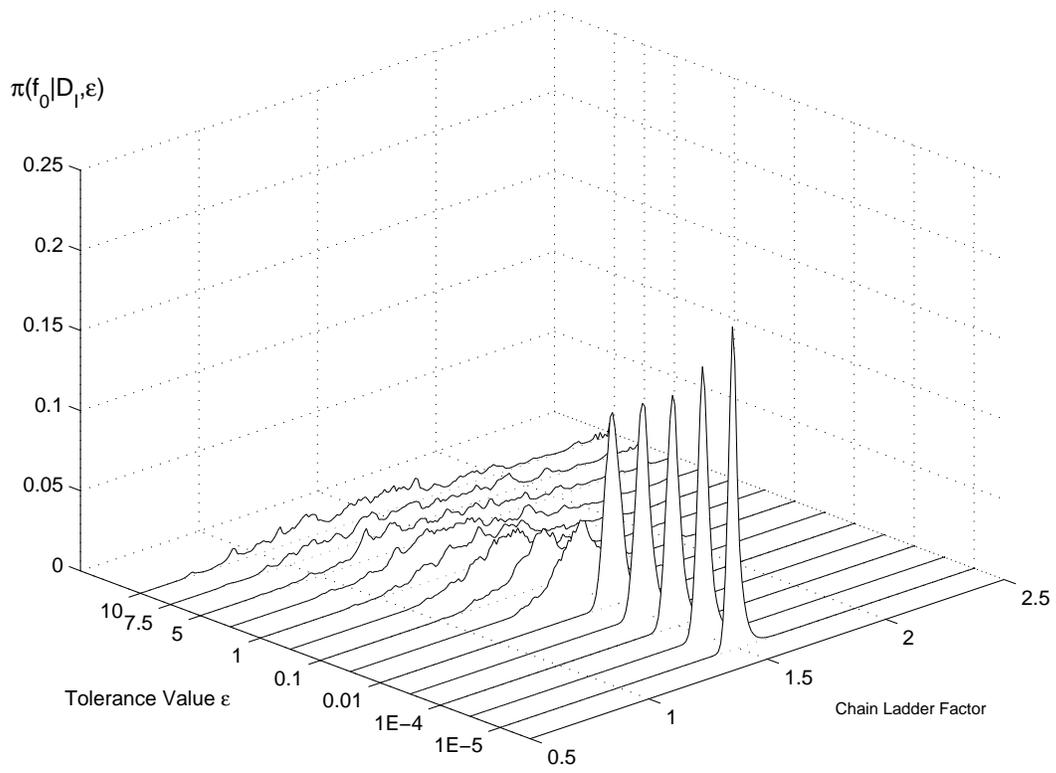}}
\caption{{}Distribution of the chain ladder factor $F_0$ as a
function of tolerance.} \label{fig6}
\end{figure}

\textbf{Dependence on tolerance $\epsilon$} \\
Figure \ref{fig6} presents a study of the histogram estimate
of the marginal posterior distribution for chain ladder factor
$\pi_{ABC}\left(f_0|\mathcal{D}_I,\epsilon^{\min}\right)$. The plot
was obtained by sampling from the full posterior
$\pi_{ABC}\left(\bm{f},\bm{\sigma}|\mathcal{D}_I,\epsilon^{\min}\right)$
for each specified tolerance value, $\epsilon^{\min}$. Then the
samples for the particular chain ladder parameter in each plot are
turned into a smoothed histogram estimate for each $\epsilon^{\min}$ and
plotted. The results of this analysis demonstrated that when
$\epsilon$ is large, in this model greater than around
$\epsilon^{\min} = 0.1$, the likelihood is not having an influence
on the ABC posterior distribution. Hence, under an MCMC-ABC
algorithm, this results in acceptance probabilities for the chain
being artificially high, resulting in estimates of the posterior
which reflect the prior distribution used (in this case a vague
prior). As $\epsilon^{\min}$ is reduced, we notice that the changes
in the estimate of the posterior distribution also reduces. The
aim of this study is to demonstrate that once $\epsilon^{\min}$
reaches a small enough level, the effect of reducing it further is
minimal on the posterior distribution. We see that changing
$\epsilon^{\min}$ from $10^{-4}$ to $10^{-5}$ has not had a
material impact on the posterior mean or variance, the change is
less than $10\%$. As a result, reducing $\epsilon^{\min}$ past this
point cannot be justified relative to the significant increase in
computational effort required to achieve such a further reduction
in $\epsilon^{\min}$.

Ultimately, we would like an algorithm which could work well for
any $\epsilon^{\min}$, the smaller the better. However, we note
that with a decreasing $\epsilon^{\min}$ in the sampler we present
in this paper, one must take additional care to ensure the Markov
chain is still mixing and not ``stuck'' in a particular state, as is
observed to be the case in all MCMC-ABC algorithms. To avoid this
acknowledged difficulty with MCMC-ABC, one should
run much longer MCMC chains or alternatively use of more
sophisticated sampling algorithms such as SMC Samplers PRC-ABC
based algorithms; see Sisson et al. ~\cite{Sisson1a}.

The conclusion of these findings is that a value of
$\epsilon^{\min}=10^{-5}$, which was used for the analysis of the
data in this paper, is suitable numerically and computationally.\\
\\
\\
\\
\\
\textbf{VaR and MSEP.} \\
In Table \ref{tab6} we present the predictive VaR at $95\%$ and
$99\%$ levels for the ultimate predicted claims, obtained from the
MCMC-ABC algorithm. These are easily obtained under the Bayesian
setting, using the MCMC-ABC posterior samples to
explicitly obtain samples from the full predictive distribution of
the cumulative claims after integrating out the parameter
uncertainty numerically. In addition to this, we present the
analysis of the MSEP under the bootstrap frequentist procedure and
the Bayesian MCMC-ABC and credibility estimates for the total
predicted cumulative claims for each accident year $i$. We also
present results for the sum of the total cumulative claims for
each accident year, and the associated parameter uncertainty and
process variance (see Section \ref{section:Bootstrap DFCL} for
details).

We can make the following conclusions from these results:
\begin{enumerate}
\item{The estimates of process variance for each $C_{i,J}$
demonstrate that the frequentist bootstrap and the credibility
estimates are very close for all accident years $i$. The Bayesian
results compare favorably with the credibility results.}
\item{The results for the parameter estimation error for the predicted cumulative claims $C_{i,J}$
demonstrate for small $i$ that the Bayesian approach results in a
smaller estimation error compared to the frequentist approach. For
large $i$, the Bayesian approach produces larger estimation error
relative to the credibility approach.}
\item{The total results for the process variance for $C=\sum_i C_{i,J}$
demonstrate that the frequentist and credibility results are very
close. Additionally, Bayesian total results are largest followed
by credibility and then frequentist estimates which is in
agreement with theoretical bounds.}
\item{The total results for the parameter estimation error for $C=\sum_i C_{i,J}$ demonstrate that
frequentist unconditional bootstrap procedure results in the
lowest total error. The Bayesian approach and credibility total
parameter errors are close. Additionally, we note that the results
in Table 7.1 of W\"{u}thrich-Merz \cite{Wuthrich}, for the total
parameter estimation error under an unconditional frequentist
bootstrap with unscaled residuals is also very close to the total
obtained under the frequentist approach.}
\end{enumerate}

\section{Discussion}
This paper has presented a distribution-free claims reserving
model under a Bayesian paradigm. A novel advanced MCMC-ABC
algorithm was developed to obtain estimates from the resulting
intractable posterior distribution of the chain ladder factors and
chain ladder variances. We assessed several aspects of this
algorithm, including the properties of the convergence of the MCMC
algorithm as a function of the distance metric approximation in
the ABC component. The methodologies performance was demonstrated
on a synthetic data set generated from known parameters. Next, it
was applied to a real claims reserving data set. The results we
obtained for predicted cumulative ultimate claims were compared to
those obtained via classical chain ladder methods and via
credibility theory. This clearly demonstrated that the algorithm
is working accurately and provides us not only with the ability to
obtain point estimates for the first and second moments of the
ultimate cumulative claims, but also with an accurate empirical
approximation of the entire distribution of the ultimate claims.
This is valuable for many reasons, including prediction of
reserves which are not based on centrality measures such as the
tail based VaR results we present.

\vspace{0.3cm} \noindent \textbf{{\large
{Acknowledgements}}}\newline \noindent The fist author thanks ETH
FIM and ETH Risk Lab for their generous financial assistance
whilst completing aspects of this work at ETH. The first author
also thanks the Department of Mathematics and Statistics at the
University of NSW for support through an Australian Postgraduate
Award and to CSIRO for support through a postgraduate research top
up scholarship. Finally, this material was based upon work
partially supported by the National Science Foundation under Grant
DMS-0635449 to the Statistical and Applied Mathematical Sciences
Institute, North Carolina, USA. Any opinions, findings, and
conclusions or recommendations expressed in this material are
those of the author(s) and do not necessarily reflect the views of
the National Science Foundation


\newpage

\begin{appendix}
\section{ABC algorithm}
\label{section: Justification for ABC}

The ABC algorithm is typically justified in the simple rejection
sampling framework. This then extends in a straightforward manner
to other sampling frameworks such as the MCMC algorithm we utilise
in this paper. We denote the posterior density from which we wish
to draw samples by $\pi\left( \theta| y\right) \propto
\pi\left(y|\theta\right)\pi\left(\theta\right)$ with $\theta \in
\Omega$, where $\Omega$ denotes support of the posterior
distribution and ${\cal Y}$ is the support for $y$.

The ABC method aims to draw from this posterior density $\pi\left(
\theta| y\right)$ without the requirement of evaluating the
computationally expensive or in our setting intractable likelihood
$\pi\left(y|\theta\right)$. The cost of avoiding this calculation
is that we obtain an ``approximation''.

~

{\bf 1st case.} We assume that the support ${\cal Y}$ is discrete.
Given an observation $y \in {\cal Y}$, we would like to sample
from $\pi\left(\theta| y\right)$. Then the original rejection
sampling algorithm reads as follows:

{\bf{Rejection Sampling ABC}}
\begin{enumerate}
\item{Sample $\theta'$ from prior $\pi\left(\theta\right)$;}
\item{Simulate synthetic data set of auxiliary variables $x|\theta' \sim \pi\left(x|\theta'\right)$;}
\item{ABC Rejection condition: if $x = y$ then accept sample $\theta'$, else reject sample and return to step 1.}
\end{enumerate}
Then the chosen $\theta'$ is distributed from $\pi(\theta|y)$. This follows from a simple rejection
argument, Denote $\{x=y\}$ if $\theta'$ was chosen. Then, the
joint density of $(\theta',x)$ conditional on $\{y, x=y\}$ is
given by
\begin{equation}
\pi(\theta,
x|y,x=y)=\frac{\pi(\theta)\pi(x|\theta)\mathbb{I}\{y\}(x)} {\int
\pi(\theta)\pi(y|\theta)d\theta} =\left\{
\begin{array}{ll}
\frac{\pi(\theta, y)}
{\pi(y)}=\pi(\theta|y)& \text{ if } x=y,\\
0&\text{ otherwise.}
\end{array}
\right.
\end{equation}
This implies that
\begin{equation}
\sum_{x\in {\cal Y}}\pi(\theta, x|y,x=y)=\pi(\theta|y).
\end{equation}
Henceforth, this algorithm generates samples $\theta^{(t)} \sim
\pi(\theta|y)$, for $t = 1,\ldots,T$.

~

{\bf 2nd case.} For more general supports ${\cal Y}$ one replaces
the strict equality $x=y$ with a tolerance $\epsilon>0$ and a
measure of discrepancy or a distance metric $\rho(x,y) \leq
\epsilon$. In this case the posterior distribution is given by
$\pi(\theta,x|y, \rho(x,y)<\epsilon)$. Implementing this algorithm
in a rejection sampling framework gives the following:

{\bf{Rejection Sampling ABC}}
\begin{enumerate}
\item{Sample $\theta'$ from prior $\pi\left(\theta\right)$;}
\item{Simulate synthetic data set of auxiliary variables $x|\theta' \sim \pi\left(x|\theta'\right)$;}
\item{ABC Rejection Condition 2: If $\rho(x,y) < \epsilon$ then accept sample $\theta'$, else reject sample and return to step 1.}
\end{enumerate}
In this case the joint density of $(\theta',x)$, conditional on
$\{y, \rho(x,y)<\epsilon\}$, is given by
\begin{equation}\label{approx2}
\pi(\theta,x|y,\rho(x,y)<\epsilon) =
\frac{\pi(\theta)~\pi(x|\theta)~\mathbb{I}\{\rho(x,y) <
\epsilon\}(x)}{\int
\pi(\theta)~\pi(x|\theta)~\mathbb{I}\{\rho(x,y) < \epsilon\}(x)~dx
d\theta }.
\end{equation}
Note that for appropriate choices of the distance metric $\rho$
and assuming the necessary continuity properties for the densities
we obtain that
\begin{equation}\label{approx3}
\lim_{\epsilon \to 0} \int_{\cal
Y}\pi(\theta,x|y,\rho(x,y)<\epsilon) dx = \pi(\theta|y).
\end{equation}
This concept was taken further with the intention of improving the
simulation efficiency by reducing the number of rejected samples.
To achieve this, sufficient statistics were used to replace the
comparison between the auxiliary variables (``synthetic data'') $x$
and the observations $y$. Denoting the sufficient statistics by
$S(y)$ and $S(x)$, allows one to decompose the
likelihood under the Fisher-Neyman factorization theorem into
$\pi(y|\theta) = f(y)g(S(y)|\theta)$ for appropriate functions $f$
and $g$. In the ABC context presented above, the consequence of
this decomposition is that when $\rho(S(y),S(x)) < \epsilon$ the 
obtained samples are from the posterior density $\pi(\theta,x|y,
\rho(S(y),S(x))<\epsilon)$ similar to \eqref{approx2}. In general,
summary statistics will be used when sufficient statistics are not
attainable.

\section{MCMC-ABC to sample from $\protect\pi_{ABC} %
\left( \bm{f},\bm{\sigma}|\mathcal{D}_{I}\right) $}

We develop an MCMC-ABC algorithm which has an adaptive proposal
mechanism and annealing of the tolerance during burn-in of the
Markov chain. Having reached the final tolerance post annealing,
denoted $\epsilon ^{\min }$, we utilise the remaining burn-in
samples to tune the proposal distribution to ensure an acceptance
probability between the range of 0.3 and 0.5 is achieved. The
optimal acceptance probability when posterior parameters are
i.i.d.~Gaussian was proven to be at 0.234; see Roberts et
al.~\cite{Gelman97}. Though our problem does not match the
required conditions for this proof, it provides a practical guide.
To achieve this, we tune the coefficient of variation of the
proposal, in our case it is the shape parameter of the Gamma proposal
distribution. We impose an additional constraint that the minimum
shape parameter value is set at $\gamma_j^{\min }$ for $j \in
\{1,\ldots,2J\}$.

\noindent \hrulefill

\textbf{MCMC-ABC algorithm using bootstrap samples.}

\begin{enumerate}
\item{For $t=0$ initialize the parameter
vector randomly, this gives $\theta _{1:2J}^{\left( 0\right)
}=\left( f_{0:J-1}^{\left( 0\right) },\sigma _{0:J-1}^{\left(
0\right) }\right) $. Initialize the proposal shape parameters
$\gamma_j \ge \gamma_j^{\min}$ for all $j \in \{1,\ldots,2J\}$.}
\item{For $t=1,\ldots,T$
\begin{enumerate}
\item{ Set $\left( \theta _{1:2J}^{\left( t\right) }\right) =\left( \theta
_{1:2J}^{\left( t-1\right) }\right) $.}
\item{ For $j=1,\ldots,2J$}
\begin{enumerate}
\item{Sample proposal $\theta _{j}^{\ast }$ from a $\Gamma
(\gamma_j, \theta_j^{(t)}/\gamma_j)$-distribution. We denote the
Gamma proposal density by $K\left( \theta _{j}^{\ast };\gamma_j
,{\theta _{j}^{(t)}}/{\gamma_j }\right)$. This gives proposed
parameter vector $ \bm{\theta}^\ast=\left( \theta
_{1:j-1}^{(t)},\theta _{j}^{\ast },\theta _{j+1:2J}^{(t)}\right)
$.}

\item{ Conditional on $\bm{\theta}^\ast=\left( \theta
_{1:j-1}^{(t)},\theta _{j}^{\ast },\theta _{j+1:2J}^{(t)}\right)
$, generate synthetic bootstrap data set $\mathcal{D}_{I}^\ast
=\mathcal{D}_{I}^\ast\left(\bm{\theta}^\ast\right)$ using the
bootstrap procedure detailed in Section \ref{section:Bootstrap
DFCL} where we replace the CL parameter estimates
$(\widehat{\bm{f}}^{(CL) },\widehat{\bm{\sigma}}^{(CL)}) $ by the
parameters $\bm{\theta}^\ast $.}

\item{ Evaluate summary statistics $S\left( \mathcal{D}_{I};0,1\right) $ and $%
S\left( \mathcal{D}_{I}^{\ast };{\mu}^\ast;{s}^\ast\right) $ and
corresponding decision function $g({\cal D}_I|{\cal D}_I^\ast)$ as
described in Section 5.}

\item{ Accept proposal with ABC acceptance probability
\begin{equation*}
A \left( \theta _{1:2J}^{\left( t\right) },\bm{\theta}^\ast
\right) =\min \left\{ 1,\frac{\pi\left( \theta _{j}^{\ast }\right)
K\left( \theta _{j}^{(t)};\gamma_{j} ,{\theta _{j}^{\ast
}}/{\gamma_{j} }\right) }{\pi\left( \theta
_{j}^{\left( t\right) }\right) K\left( \theta _{j}^{\ast };\gamma_{j} ,{%
\theta _{j}^{(t)}}/{\gamma_{j} }\right) } ~g({\cal D}_I|{\cal
D}_I^\ast)
 \right\}.
\end{equation*}
That is, simulate $U\sim {\cal U}(0,1)$ and set $\theta
_{j}^{\left( t\right) }=\theta _{j}^{\ast }$ if $U<A \left( \theta
_{1:2J}^{\left( t\right) },\bm{\theta}^\ast \right)$. }

\item {If $100 \le t \le T_{b}$ and $\epsilon_t = \epsilon^{\text{min}}$
then check to see if tuning of the proposal is required. Define the
average acceptance probability over the last 100 iterations of
updates for parameter $i$ by $\bar{a}_i^{(t-100:t)}$ and consider
the adaption:
\begin{eqnarray*}
 \gamma_j^* = \left\{ \begin{array}{l l l} 0.9\gamma_j & \quad
\mbox{ if $\bar{a}_i^{(t-100:t)}<0.3$ and $\gamma_j >
\gamma_j^{\text{min}}$},\\
1.1\gamma_j & \quad \mbox{ if $\bar{a}_i^{(t-100:t)}>0.5$},\\
\gamma_j & \quad \mbox{ otherwise.}\\
\end{array}
\right.
\end{eqnarray*}
Then set the proposal shape parameter as $\gamma_j = \text{max}\{\gamma_j^*, \gamma_j^{\text{min}}\}$.
}
\end{enumerate}
\end{enumerate}
}
\end{enumerate}

\noindent\hrulefill

The MCMC-ABC algorithm presented can be enhanced by utilising an
idea of Gramacy et al. \cite{Gramacy} in an ABC setting. This
involves a combination of tempering the tolerance $\{\epsilon
_{t}\}_{t=1:T}$ and importance sampling corrections.

\subsection{ABC\ algorithmic choices for the time series DFCL\
model} \label{section:ABC Settings for DFCL model}

We start with the choices of the ABC components.

\begin{itemize}
\item \textbf{Generation of a synthetic data set}:
\label{synthetic data} Note that in this setting not only is the
likelihood intractable but also the generation of a
synthetic data set $\mathcal{D}_{I}^{\ast }$ given the current
parameter values $\bm{F},\bm{\Xi}$ is not straightforward. The
synthetic data set $\mathcal{D}_{I}^{\ast }$ is generated using
the bootstrap procedure described in Section
\ref{section:Bootstrap DFCL}. Note that both the bootstrap
residual $\widetilde{\varepsilon}_{i,j}$ and the bootstrap samples
$\mathcal{D}_I^\ast$ are functions of the parameter choices; see Section 4.1. Therefore we generate for given
$\bm{F}=\bm{f}$ and $\bm{\Xi}=\bm{\sigma}$ the bootstrap residuals
$\widetilde{\varepsilon}_{i,j}=
\widetilde{\varepsilon}_{i,j}(f_{j-1},\sigma_{j-1})$ and the
bootstrap samples
$\mathcal{D}_I^\ast=\mathcal{D}_I^\ast(\bm{f},\bm{\sigma})$
according to the non-parametric bootstrap (see Section 4.1) where we replace the CL parameter estimates
$(\widehat{\bm{f}}^{(CL) },\widehat{\bm{\sigma}}^{(CL)}) $ by the
parameters $\theta=(\bm{F},\bm{\Xi})$.

\item \textbf{Summary statistics}: We introduce summary statistics to replace sufficient statistics when they
are not attainable for a given model. Then, in order to define the
decision function $g$, we introduce summary statistics; see
Appendix \ref{section: Justification for ABC}. For the observed
data $\mathcal{D}_I$ we define the vector
\begin{eqnarray*}
S\left( \mathcal{D}_{I}; 0, 1\right)  &=&\left( S_{1},\ldots,S_{n+2}\right)  \\
&=&\left(C_{0,1},\ldots,C_{0,J}, C_{1,1},\ldots,C_{0,J-1}, \ldots,
C_{I-2,1},C_{I-2,2}, C_{I-1,1};0, 1 \right),
\end{eqnarray*}
where $n$ denotes the number of residuals
$\widetilde{\varepsilon}_{i,j}$. For given
$\theta=(\bm{F},\bm{\Xi})$, we generate the bootstrap sample
$\mathcal{D}_I^\ast= \mathcal{D}_I^\ast(\bm{F},\bm{\Xi})$ as
described above. The corresponding residuals
$\widetilde{\varepsilon}_{i,j}=
\widetilde{\varepsilon}_{i,j}(F_{j-1},\Xi_{j-1})$ should also be
close to the standardized observations. Therefore, we define its
empirical mean and standard deviation by
\begin{eqnarray}\label{mu1}
\mu^\ast&=&\mu^\ast(\bm{F},\bm{\Xi}) ~=~\frac{1}{n}
\sum_{i,j}\widetilde{\varepsilon}_{i,j}(F_{j-1},\Xi_{j-1}),
\\\label{mu2}
s^\ast&=&s^\ast(\bm{F},\bm{\Xi}) ~=~\left[\frac{1}{n-1} \sum_{i,j}
\left(\widetilde{\varepsilon}_{i,j}(F_{j-1},\Xi_{j-1})-
\mu^\ast(\bm{F},\bm{\Xi})\right)^2\right]^{1/2}.
\end{eqnarray}
Hence, the summary statistics for the synthetic data is given by
\begin{equation*}
S\left( \mathcal{D}^\ast_{I}; \mu^\ast, s^\ast\right)
=\left(C^\ast_{0,1},\ldots,C^\ast_{0,J},
C^\ast_{1,1},\ldots,C^\ast_{0,J-1}, \ldots,
C^\ast_{I-2,1},C^\ast_{I-2,2}, C^\ast_{I-1,1};\mu^\ast, s^\ast
\right).
\end{equation*}

\item \textbf{Distance metrics}:

\begin{itemize}
\item \textit{Mahlanobis distance and scaled Euclidean distance}

Here we draw on the analysis of Sisson et al.~\cite{Fan} that
proposes the use of the Mahlanobis distance metric given by%
\begin{eqnarray*}
&&\hspace{-1cm} \rho \left( S\left( \mathcal{D}_{I};0,1\right)
,S\left( \mathcal{D}_{I}
^{\ast}; \mu^\ast, s^\ast\right) \right) \\
&=& \left[ S\left( \mathcal{D}_{I};0,1\right) -S \left(
\mathcal{D}_{I} ^{\ast}; \mu^\ast, s^\ast\right) \right] ^{\top }
~\Sigma _{\mathcal{D}_{I}}^{-1}~ \left[ S\left(
\mathcal{D}_{I};0,1\right) -S \left( \mathcal{D}_{I} ^{\ast};
\mu^\ast, s^\ast\right) \right] ,
\end{eqnarray*}%
where the covariance matrix $\Sigma _{\mathcal{D}_{I}} $ is an
appropriate scaling described in Appendix \ref{section: Mahlanobis
Distance Metric}. The scaled Euclidean distance is obtained when
we only consider the diagonal elements of the covariance matrix
$\Sigma _{\mathcal{D}_{I}}$.

Note, the covariance matrix $\Sigma _{\mathcal{D}_{I}}$ provides a
weighting on each element of the vector of summary statistics to
ensure they are scaled appropriately according to their influence
on the ABC approximation. There are many other such weighting
schemes one could conceive.

\item \textit{Manhattan ``City Block'' distance}

We consider the $L^1$-distance given by
\begin{equation*}
\rho \left( S\left( \mathcal{D}_{I};0,1\right) ,S\left(
\mathcal{D}_{I} ^{\ast}; \mu^\ast, s^\ast\right) \right)
=\sum_{i=1}^{n+2}\left| S_{i}\left( \mathcal{D}%
_{I};0,1\right) -S_{i}\left( \mathcal{D}_{I} ^{\ast}; \mu^\ast,
s^\ast\right) \right|.
\end{equation*}
\end{itemize}

\item \textbf{Decision function}: We work with a hard
decision function given by
\begin{equation*}
g\left( \mathcal{D}_{I}| \mathcal{D}_I^\ast \right)
=\mathbb{I}\left\{ \rho \left( S\left( \mathcal{D}_{I};0,1\right)
,S\left( \mathcal{D}_{I} ^{\ast}; \mu^\ast, s^\ast\right) \right)
<\epsilon \right\} .
\end{equation*}

\item \textbf{Tolerance schedule}: We
use the sequence
\begin{equation*}
\epsilon _{t}=\max \left\{ 20,000-10t,\epsilon^{\min } \right\} .
\end{equation*}%
Note, the use of an MCMC-ABC algorithm can result in ``sticking'' of
the chain for extended periods. Therefore, one should carefully
monitor convergence diagnostics of the resulting Markov chain for
a given tolerance schedule. There is a trade-off between the
length of the Markov chain required for samples approximately from
the stationary distribution and the bias introduced by non zero
tolerance. In this paper we set $\epsilon^{\min }$ via preliminary
analysis of the Markov chain sampler mixing rates for a transition
kernel with coefficient of variation set to one.

We note that in general, practitioners will have a required
precision in posterior estimates that can be directly
used to determine, for a given computational budget, a suitable
tolerance $\epsilon^{\min }$.

\item \textbf{Convergence diagnostics:}
We stress that when using an MCMC-ABC algorithm, it is crucial to
carefully monitor the convergence diagnostics of the Markov chain.
This is more important in the ABC context than in the general MCMC
context due to the possibility of extended rejections where the
Markov chain can stick in a given state for long periods. This can
be combatted in several ways which will be discussed once the
algorithm is presented.

The convergence diagnostics we consider are evaluated only on
samples post annealing of the tolerance threshold and after an
initial burn-in period once tolerance of $\epsilon^{\min }$ is
reached. If the total chain has length $T$, the initial burn-in
stage will correspond to the first $T_{b}$ samples and we define
$\widetilde{T}=T-T_{b}$. We denote by
$\{\theta_i^{(t)}\}_{t=1:\widetilde{T}}$ the Markov chain of the
$i$-th parameter after burn-in. The diagnostics we consider are
given by:
\begin{itemize}
\item{\textit{Autocorrelation.}
This convergence diagnostic will monitor serial correlation in the
Markov chain. For given Markov chain samples for the $i$-th
parameter $\{\theta_i^{(t)}\}_{t=1:\widetilde{T}}$, we define the
biased autocorrelation estimate at lag $\tau$ by
\begin{equation}
\widehat{ACF}(\theta_i,\tau) =
\frac{1}{(\widetilde{T}-\tau)\hat{\sigma}\left(\theta_i\right)}\sum_{t=1}^{\widetilde{T}-\tau}[\theta_i^{(t)}-\widehat{\mu}\left(\theta_i\right)][\theta_i^{(t+\tau)}-\widehat{\mu}\left(\theta_i\right)],
\end{equation}
where $\widehat{\mu}\left(\theta_i\right)$ and
$\hat{\sigma}\left(\theta_i\right)$ are the estimated mean and
standard deviation of $\theta_i$.
\item{ \textit{Geweke \cite{Geweke1991} time series diagnostic.}}
For parameter $\theta_i$ it is calculated as follows:
\begin{enumerate}
\item{Split the Markov chain samples into two sequences, $\{\theta_i^{(t)}\}_{t=1:T_1}$ and
$\{\theta_i^{(t)}\}_{t=T^*:\widetilde{T}}$, such that $T^* = \widetilde{T}
- T_2 + 1$, and with ratios $T_1/\widetilde{T}$ and $T_2/\widetilde{T}$
fixed such that $(T_1+T_2)/\widetilde{T} < 1$ for all $\widetilde{T}$.}
\item{Evaluate $\widehat{\mu}\left(\theta_i^{T_1}\right)$ and $\widehat{\mu}\left(\theta_i^{T_2}\right)$ corresponding to the sample means on each sub sequence. }
\item{ Evaluate consistent spectral density estimates for each sub sequence, at frequency 0, denoted $\widehat{SD}(0;T_1,\theta_i)$ and
$\widehat{SD}(0;T_2,\theta_i)$. The spectral density estimator
considered in this paper is the classical non-parametric
periodogram or power spectral density estimator. We use Welch's
method with a Hanning window; for details see Appendix
\ref{section: Power Spectral Density Estimate}.
 }
\item{Evaluate convergence diagnostic given by \newline
$Z_{\widetilde{T}}=\frac{\widehat{\mu}\left(\theta_i^{T_1}\right)-\widehat{\mu}\left(\theta_i^{T_2}\right)}{T_1^{-1}\widehat{SD}(0;T_1,\theta_i)+
T_2^{-1}\widehat{SD}(0;T_2,\theta_i) }.$ \\
According to the central limit theorem, as $\widetilde{T} \rightarrow \infty$ one has that $Z_{\widetilde{T}} \rightarrow {\cal
N}(0,1)$ if the sequence $\{\theta_i^{(t)}\}_{t=1:\widetilde{T}}$ is
stationary.}
\end{enumerate}
}
\item{ \textit{Gelman-Rubin \cite{Gelman1a} R-statistic diagnostic.}
This approach to convergence analysis requires that one runs
multiple parallel independent Markov chains each starting at
randomly selected initial starting points (we run five chains). For
comparison purposes we split the total computational budget of
$\widetilde{T}$ into $T_1=T_2=\ldots=T_5=\frac{\widetilde{T}}{5}$. The convergence diagnostic for parameter $\theta_i$  is calculated using the following steps:
\begin{enumerate}
\item{Generate five independent Markov chain sequences, producing the chains for parameter $\theta_i$ denoted $\{\theta_{i,k}^{(t)}\}_{t=1:T_k}$ for $k \in \{1,\ldots,5\}$.}
\item{Calculate the sample means $\widehat{\mu}\left(\theta_i^{T_k}\right)$ for each sequence and the overall mean $\widehat{\mu}\left(\theta_i^{\widetilde{T}}\right)$.}
\item{Calculate the variance of the sequence means \newline $\frac{1}{4}\sum_{k=1}^{5}\left(\widehat{\mu}\left(\theta_i^{T_k}\right)-\widehat{\mu}\left(\theta_i^{\widetilde{T}}\right)\right)^2\stackrel{def.}{=}B_i/T_k .$}
\item{Calculate the within-sequence variances $\widehat{s}^2\left(\theta_i^{T_k}\right)$ for each sequence.}
\item{Calculate the average within-sequence variance, $\frac{1}{5}\sum_{k=1}^{5}\widehat{s}^2\left(\theta_i^{T_k}\right)\stackrel{def.}{=}W_i$.}
\item{Estimate the target posterior variance for parameter $\theta_i$ by the weighted linear combination
$\widehat{\sigma}^2\left(\theta_i^{\widetilde{T}}\right) =
\frac{T_k-1}{T_k}W_i + \frac{1}{T_k}B_i$. This estimate is
unbiased for samples which are from the stationary distribution.
In the case in which not all sub chains have reached stationarity,
this overestimates the posterior variance for a finite $\widetilde{T}$
but asymptotically, $\widetilde{T} \rightarrow \infty $, it converges
to the posterior variance.}
\item{Improve on the Gaussian estimate of the target posterior given
by\\
$\mathcal{N}(\widehat{\mu}\left(\theta_i^{\widetilde{T}}\right),\widehat{\sigma}^2\left(\theta_i^{\widetilde{T}}\right))$
by accounting for sampling variability in the estimates of the
posterior mean and variance. This can be achieved by making a
Student-t approximation with location
$\widehat{\mu}\left(\theta_i^{\widetilde{T}}\right)$, scale
$\sqrt{\widehat{V}_i}$ and degrees of freedom $df_i$, each given
respectively by: \newline
$\widehat{V}_i=\widehat{\sigma}^2\left(\theta_i^{\widetilde{T}}\right)+\frac{B_i}{\widetilde{T}}$
and $df_i =
\frac{2\widehat{V}_i^2}{\widehat{\text{Var}}(\widehat{V}_i)}$,
 where the variance is estimated as\\
\begin{equation}
\begin{split}
\widehat{\text{Var}}\left(\widehat{V}_i\right) &=
\frac{1}{5}\left(\frac{T_1-1}{T_1}\right)^2\widehat{\text{Var}}\left(\widehat{s}^2\left(\theta_i^{T_k}\right)\right)+
\left(\frac{6}{\sqrt{2}\widetilde{T}}\right)^2B_i^2 \\
&+\frac{12(T_1-1)}{25T_1}\widehat{\text{Cov}}\left(\widehat{s}^2\left(\theta_i^{T_k}\right),\widehat{\mu}\left(\theta_i^{\widetilde{T}}\right)\right) \\
&-\frac{24(T_1-1)}{25T_1}\widehat{\mu}\left(\theta_i^{\widetilde{T}}\right)\widehat{\text{Cov}}\left(\widehat{s}^2\left(\theta_i^{T_k}\right),\widehat{\mu}\left(\theta_i^{\widetilde{T}}\right)\right).
\end{split}
\end{equation}
Note, the covariance terms are estimated empirically using the
within sequence estimates of the mean and variance obtained for
each sequence.}
\item{Calculate the convergence diagnostic $\sqrt{\widehat{R}} =
\sqrt{\frac{\widehat{V}_i df_i}{W_i\left(df_i-2\right)}}$, where
as $\widetilde{T} \rightarrow \infty$ one can prove that $\widehat{R}
\rightarrow 1$. This convergence diagnostic monitors the scale
factor by which the current distribution for $\theta_i$ may be
reduced if simulations are continued for $\widetilde{T} \rightarrow
\infty$.}
\end{enumerate}

 }
\end{itemize}
\end{itemize}

\section{Scaling of statistics in distance metrics}
\label{section: Mahlanobis Distance Metric}

In the Mahlanobis distance metric, estimation of the scaling
weights is given by the covariance  $\Sigma _{\mathcal{D}_{I}}
=\mathrm{Cov}\left( \left. S\left( \mathcal{D}_{I}^{\ast
};\widetilde{\mu}, \widetilde{s}\right) \right\vert
\mathcal{D}_{I}\right) $, where $\widetilde{\mu}$ and
$\widetilde{s}$ are the sample mean and standard deviation of $n$
i.i.d.~residuals $\varepsilon_{i,j}$ (see also
\eqref{mu1}-\eqref{mu2}). Next we outline the estimation of
${\Sigma} _{\mathcal{D}_{I}}$ by a matrix
$\widehat{\Sigma}^{CL}_{\mathcal{D}_{I}}$.

\begin{itemize}
\item {Starting with the elements $
\widehat{\Sigma}^{CL}_{\mathcal{D}_{I}}(k,l)$ with $k,l \in
\{1,\ldots,n\}$,
 we obtain from the conditional resampling bootstrap
\begin{itemize}

\item{$\mathrm{Cov}\left(\left. C^\ast_{i,j}, C^\ast_{i',j'} \right\vert
{\cal D}_I,\widehat{\bm{f}}^{(CL)}, \widehat{\bm{\sigma}}^{(CL)}
\right) = 0$ if $i\neq i'$ or $j \neq j'$}

\item{$
\mathrm{Var}\left(\left. C^\ast_{i,j}\right\vert {\cal
D}_I,\widehat{\bm{f}}^{(CL)}, \widehat{\bm{\sigma}}^{(CL)} \right)
=\widehat{\sigma}_{j-1}^{2(CL)}C_{i,j-1}.$ }
\end{itemize}
}

\item {Considering the elements $k \in \{n+1,n+2\}, l \in
\{1,\ldots,n\}$ and also $k \in \{1,\ldots,n\}, l \in \{n+1,n+2\}$
of the covariance matrix $\Sigma_{{\cal D}_I}$, for simplicity we
set $\widehat{\Sigma}^{CL}_{\mathcal{D}_{I}}(k,l) = 0$. }

\item {Considering elements $k,l \in \{n+1,n+2\}$,
we assess now $\mathrm{Cov}(\widetilde{\mu}, \widetilde{s})$ either
analytically or numerically by simulation of appropriate
i.i.d.~residuals.


\textbf{Parametric Approximation}
\begin{itemize}
\item{In approximating $\tilde{\mu}$ and $\tilde{s}$ we assume i.i.d. samples $\varepsilon_{i,j} {\sim}
\mathcal{N}\left(0,1\right)$.}
\item{Using the assumptions we know that: \\
$\mathrm{Var}(\widetilde{\mu})=\frac{1}{n}$,\\
$\mathrm{Var}(\widetilde{s})=\frac{1}{(n-1)^2}\left[\left(1+\frac{4}{n^2}+\frac{1}{n^2}\right)\sum_{s=1}^{n}\mathrm{Var}\left(\widetilde{\varepsilon}^2_{s}\right)\right]
 =\frac{1}{(n-1)^2}[2n(1+\frac{5}{n^2})]$, \\
 $\mathrm{Cov}(\widetilde{\mu},\widetilde{s}) =
 \frac{1}{2(n-1)^2}[1-\frac{2}{n}]$.
}
\item{ Under these assumptions:\\
1. If the distribution of $\varepsilon_{i,j}$ is skewed then it is
more appropriate to do a numerical approximation with the observed residuals from the bootstrap algorithm.\\
2. The precision $\epsilon_t$ from the MCMC-ABC algorithm should
depend on the size of the claims triangle, that is, the number of
residuals $n$.}
\end{itemize}
}
\end{itemize}

\section{Estimating the Spectral Density}
\label{section: Power Spectral Density Estimate}

This is calculated via a modified technique using Welch's method;
see Proakis-Manolakis \cite{Proakis1}, 910-913 . This
involves performing the following steps:
\begin{itemize}
\item{Split each sequence $\{\theta_i^{(t)}\}_{t=1:T_1}$ and
$\{\theta_i^{(t)}\}_{t=T^*:\widetilde{T}}$ into $L=20$
non-overlapping blocks of length $N$.}
\item{Apply a Hanning window function
$w(t)=0.5\left(1-\text{cos}\left(\frac{2 \pi
t}{N-1}\right)\right)$ to the samples of the Markov chain in each
block.}
\item{Take the discrete Fourier transform (DFT) of
each windowed block given by $\Tilde{\Theta}_i^l(k) =
\sum_{t=0}^{N-1}\theta_i^{(t)}\text{exp}\left(-\frac{2 \pi
ikt}{N}\right)$.}
\item{Estimate the spectral density (SD) as
$\widehat{SD}(w_k)=\frac{1}{L}\sum_{l=0}^{L-1}\Tilde{\Theta}_i^{l}(k)$.
}
\end{itemize}

\end{appendix}

\newpage

{\footnotesize
\begin{table}[tbph]
\begin{center}
{\footnotesize {\scriptsize {\
\begin{tabular}{c|cccccccccc}
{Year} & {$0$} & {$1$} & {$2$} & {$3$} & {$4$} & {$5$} & {$6$} & {$7$} & {$8$%
} & {$9$} \\ \hline {$0$} & 248.97 & 299.47 & 357.00 & 418.61 &
473.63 & 563.35 & 693.22 & 796.84 & 914.95 &
\multicolumn{1}{c|}{1,084.24}
\\ \cline{11-11} {$1$} & 186.72 & 201.99 & 227.23 & 271.18 &
305.16 & 379.37 & 466.16 & 554.30 & 660.75 & \multicolumn{1}{|c|}{} \\
\cline{10-10} {$2$} & 172.58 & 207.48 & 250.37 & 304.44 & 356.92 &
417.60 & 477.99 & 542.25 & \multicolumn{1}{|c}{} & \multicolumn{1}{c|}{} \\
\cline{9-9} {$3$} & 195.19 & 229.06 & 290.83 & 320.11 & 367.60 & 469.93 & 543.40 & \multicolumn{1}{|c}{} &  & \multicolumn{1}{c|}{} \\
\cline{8-8} {$4$} & 131.00 & 168.50 & 198.18 & 219.26 & 270.00 & 344.63 & \multicolumn{1}{|c}{} &  &  & \multicolumn{1}{c|}{} \\
\cline{7-7} {$5$} & 163.58 & 181.16 & 222.10 & 246.78 & 303.00 &
\multicolumn{1}{|c}{} & &  &  & \multicolumn{1}{c|}{} \\
\cline{6-6} {$6$} & 294.30 & 373.08 & 477.16 & 566.20 &
\multicolumn{1}{|c}{} &  &  &  & & \multicolumn{1}{c|}{} \\
\cline{5-5} {$7$} & 529.31 & 577.71 & 805.95 &
\multicolumn{1}{|c}{} & & & & & &
\multicolumn{1}{c|}{} \\
\cline{4-4} {$8$} & 249.00 & 321.83 & \multicolumn{1}{|c}{} &  &
& & & &
& \multicolumn{1}{c|}{}\\
\cline{3-3} {$9$} & 140.41 & \multicolumn{1}{|c}{} &  &  &  & &  &
& & \multicolumn{1}{c|}{}
\\ \cline{2-11}\\
{$f_j$} & {$1.2$} & {$1.2$} & {$1.2$} & {$1.2$} & {$1.2$} &
{$1.2$} & {$1.2$} & {$1.2$} &
{$1.2$} & {$1.2$}\\
{$\sigma^2_j$} & {$1$} & {$1$} & {$1$} & {$1$} & {$1$} & {$1$} &
{$1$} & {$1$} & {$1$} & {$1$}
\end{tabular}
} }  }
\end{center}
\caption{Synthetic Data - Cumulative claims $C_{i,j}$ for each
accident year $i$ and development year $j,$ $i+j\leq
I$.}\label{tab2}
\end{table}
{\scriptsize {\ }}}


{\footnotesize
\begin{table}[ptbh]
\begin{center}
{\footnotesize {\scriptsize {\
\begin{tabular}{c|cccccccccc}
{Year} & {$0$} & {$1$} & {$2$} & {$3$} & {$4$} & {$5$} & {$6$} & {$7$} & {$8$%
} & {$9$} \\ \hline
{$0$} & {$594.6975$} & {$372.1236$} & {$89.5717$} & {$20.7760$} & {$20.6704 $%
} & {$6.2124$} & {$6.5813$} & {$1.4850$} & {$1.1130$} & \multicolumn{1}{c|}{$%
1.5813$} \\ \cline{11-11} {$1$} & {$634.6756$} & {$324.6406$} &
{$72.3222$} & {$15.1797$} & {$6.7824$} & {$3.6603$} & {$5.2752$} &
{$1.1186$} & {$1.1646$} & \multicolumn{1}{|c|}{}
\\ \cline{10-10}
{$2$} & {$626.9090$} & {$297.6223$} & {$84.7053$} & {$26.2768$} & {$15.2703 $%
} & {$6.5444$} & {$5.3545$} & {$0.8924$} & \multicolumn{1}{|c}{} &
\multicolumn{1}{c|}{} \\ \cline{9-9}
{$3$} & {$586.3015$} & {$268.3224$} & {$72.2532$} & {$19.0653$} & {$13.2976 $%
} & {$8.8340$} & {$4.3329$} & \multicolumn{1}{|c}{} &  &
\multicolumn{1}{c|}{ } \\ \cline{8-8}
{$4$} & {$577.8885$} & {$274.5229$} & {$65.3894$} & {$27.3395$} & {$23.0288 $%
} & {$10.5224$} & \multicolumn{1}{|c}{} &  &  & \multicolumn{1}{c|}{} \\
\cline{7-7}
{$5$} & {$618.4793$} & {$282.8338$} & {$57.2765$} & {\ $24.4899$} & {$%
10.4957 $} & \multicolumn{1}{|c}{} &  &  &  & \multicolumn{1}{c|}{} \\
\cline{6-6} {$6$} & {$560.0184$} & {$289.3207$} & {$56.3114$} & {\
$22.5517$} & \multicolumn{1}{|c}{} &  &  &  &  &
\multicolumn{1}{c|}{} \\ \cline{5-5} {$7$} & {$528.8066$} &
{$244.0103$} & {$52.8043$} & \multicolumn{1}{|c}{} & &  &  &  &  &
\multicolumn{1}{c|}{} \\ \cline{4-4} {$8$} & {$529.0793$} &
{$235.7936$} & \multicolumn{1}{|c}{} &  &  &  &  &  & &
\multicolumn{1}{c|}{} \\ \cline{3-3} {$9$} & {$567.5568$} &
\multicolumn{1}{|c}{} &  &  &  &  &  &  &  & \multicolumn{1}{c|}{}
\\ \cline{2-11}
\end{tabular}
} }  }
\end{center}
\caption{Real Data - Incremental claims
$Y_{i,j}=C_{i,j}-C_{i,j-1}$ for each accident year $i$ and
development year $j,$ $i+j\leq I$.} \label{tab3}
\end{table}
}

\newpage 
\begin{landscape}
{\footnotesize
\begin{table}[ptbh]
\begin{center}
{\footnotesize {\scriptsize {\
\begin{tabular}{|l||l|l|l|l|l|l|l|l|l|}
\hline DFCL model & $j=0$ & $j=1$ & $j=2$ & $j=3$ &
$j=4$ & $j=5$ & $j=6$ & $j=7$ & $j=8$ \\
\hline
$f_j$ & 1.20 & 1.20 & 1.20 & 1.20 & 1.20 & 1.20 & 1.20 & 1.20 & 1.20\\
\hline
$\widehat{f}^{(CL)}_{j}$ & 1.20 (2.40E-2) & 1.22 (3.27E-2) & 1.16 (2.46E-2) & 1.17 (2.44E-2) & 1.23 (2.63E-2) & 1.19 (2.78E-2) & 1.16 (2.59E-2) & 1.17 (2.10E-2) & 1.19 (2.51E-2)\\
\hline
$\widehat{f}^{(MAP)}_{j}|\sigma_{0:J-1}$ & 1.07 (0.02) & 1.19 (0.02) & 1.05 (0.02) & 1.04 (0.02) & 1.10 (0.02) & 1.08 (0.02) & 0.97 (0.02) & 1.19 (0.03) & 1.14 (0.04)\\
\hline
$\widehat{f}^{(MMSE)}_{j}|\sigma_{0:J-1}$ & 1.19 (1.34E-2) & 1.21 (1.38E-2) & 1.18 (1.27E-2) & 1.19 (1.30E-2) & 1.17 (1.37E-2) & 1.18 (1.53E-2) & 1.20 (1.60E-2) & 1.18 (1.73E-2) & 1.19 (2.35E-2)\\
\hline
$\widehat{\sigma}_{f_{j}}|\sigma_{0:J-1}$ & 0.23 (4.00E-3) & 0.22 (3.1E-3) & 0.20 (3.1E-3) & 0.21 (3.2E-3) & 0.22 (3.9E-3) & 0.27 (1.01E-2) & 0.35 (1.24E-2) & 0.44 (1.41E-2) & 0.70 (1.60E-2)\\
\hline
$[\hat{q}_{0.05},\hat{q}_{0.95}]|\sigma_{0:J-1}$ & [0.75,1.50] & [0.77,1.50] & [0.76,1.41] & [0.75,1.44] & [0.82,1.51] & [0.78,1.52] & [0.65,1.60] & [0.46,1.79] & [0.25,2.50]\\
\hline
$\widehat{f}^{(MAP)}_{j}$ & 1.15 (0.02) & 1.13 (0.02) & 1.06 (0.02) & 1.09 (0.02) & 1.15 (0.02) & 1.19 (0.02) & 1.12 (0.03) & 1.08 (0.03) & 1.06 (0.04)\\
\hline
$\widehat{f}^{(MMSE)}_{j}$ & 1.19 (0.01) & 1.18 (0.01) & 1.17 (0.01) & 1.18 (0.01) & 1.16 (0.01) & 1.20 (0.02) & 1.18 (0.03) & 1.16 (0.02) & 1.20 (0.02)\\
\hline
$\widehat{\sigma}_{f_{j}}$ & 0.24 (5.1E-3) & 0.24 (4.4E-3) & 0.23 (5.0E-3) & 0.26 (5.8E-3) & 0.25 (5.6E-3) & 0.25 (5.7E-3) & 0.40 (0.01) & 0.49 (0.02) & 0.68 (0.02) \\
\hline
$[\hat{q}_{0.05},\hat{q}_{0.95}]$ & [0.66,1.48] & [0.74,1.54] & [0.67,1.42] & [0.65,1.47] & [0.74,1.50] & [0.74,1.50] & [0.22,1.54] & [0.35,1.95] & [0.1,2.50]\\
\hline
$Ave[A\left(\theta_{1:2J},f_{j}\right)]$ & 0.21 & 0.21 & 0.19 & 0.22 & 0.25 & 0.21 & 0.22 & 0.20 & 0.24\\
\hline
$\sigma_j^2$ & 1 & 1 & 1 & 1 & 1 & 1 & 1 & 1 & 1\\
\hline
$\widehat{\sigma}^{2(CL)}_{j}$ & 1.02 (0.29) & 0.75 (1.44) & 0.51 (1.02) & 0.49 (0.91) & 0.71 (1.18) & 0.72 (1.89) & 0.25 (1.84) & 0.31 (1.40) & 0.25 (0.77)\\
\hline
$\widehat{\sigma}^{2(MAP)}_{j}$ & 0.58 (0.06) & 0.96 (0.06) & 0.54 (0.05) & 0.78 (0.05) & 0.78 (0.05) & 0.81 (0.04) & 0.61 (0.04) & 0.79 (0.04) & 0.56 (0.04)\\
\hline
$\widehat{\sigma}^{2(MMSE)}_{j}$ & 1.11 (0.03) & 1.18 (0.03) & 1.14 (0.04) & 1.31 (0.03) & 1.29 (0.03) & 1.19 (0.02) & 1.16 (0.03) & 1.14 (0.03) & 1.05 (0.02)\\
\hline
$\widehat{\sigma}_{\sigma_{j}}$ & 0.83 (0.02) & 0.79 (0.02) & 0.82 (0.02) & 0.80 (0.02) & 0.79 (0.02) & 0.72 (0.02) & 0.77 (0.02) & 0.78 (0.02) & 0.71 (0.02)\\
\hline
$[\hat{q}_{0.05},\hat{q}_{0.95}]$ & [0.33,2.89] & [0.33,2.79] & [0.25,2.91] & [0.32,2.87] & [0.33,2.82] & [0.27,2.59] & [0.21,2.66] & [0.17,2.62] & [0.22,2.42]\\
\hline
$Ave[A\left(\theta_{1:2J},\sigma_{j}\right)]$ & 0.23 & 0.24 & 0.24 & 0.23 & 0.24 & 0.24 & 0.24 & 0.24 & 0.25\\
\hline
\end{tabular}
} }  }
\end{center}
\caption{Comparison of Bayesian estimates for the chain ladder
factors and variances versus classical estimates, in the case of synthetic
data. Numerical standard errors in estimates are presented in
brackets.} \label{tab3}
\end{table}
}
\end{landscape}

\newpage

%

\begin{landscape}
{\footnotesize
\begin{table}[ptbh]
\begin{center}
{\footnotesize {\scriptsize {\
\begin{tabular}{cc|cccccccccc|c}
\hline \multicolumn{1}{c|}{Parameters} & {Year} & {$0$} & {$1$} & {$2$} & {$3$} & {$4$} & {$5$} & {$6$} & {$7$} & {$8$} & {$9$} & {$\widehat{C}_{i,J}^{(CL)}-{C}_{i,I-i}$} \\
\hline \multicolumn{1}{c|}{$\bm{f^{(CL)}}$} & {$0$} & {$$} & {$$} & {$$} & {$$} & {$$} & {$$} & {$$} & {$$} & {$$} & \multicolumn{1}{c|}{$$} & {$0$} \\
\multicolumn{1}{c|}{$\bm{f^{(MMSE)}}$} & {$$} & {$$} & {$$} & {$$} & {$$} & {$$} & {$$} & {$$} & {$$} & {$$} & \multicolumn{1}{c|}{$$} & {$0$} \\
\hline \multicolumn{1}{c|}{$\bm{f^{(CL)}}$} & {$1$} & {$$} & {$$} & {$$} & {$$} & {$$} & {$$} & {$$} & {$$} & {$$} & \multicolumn{1}{c|}{$10,663,318$} & {$15,126$} \\
\multicolumn{1}{c|}{$\bm{f^{(MMSE)}}$} & {$$} & {$$} & {$$} & {$$} & {$$} & {$$} & {$$} & {$$} & {$$} & {$$} & \multicolumn{1}{c|}{$10,663,099$} & {$14,907$} \\
\hline \multicolumn{1}{c|}{$\bm{f^{(CL)}}$} & {$2$} & {$$} & {$$} & {$$} & {$$} & {$$} & {$$} & {$$} & {$$} & {$10,646,884$} & \multicolumn{1}{c|}{$10,662,008$} & {$26,257$} \\
\multicolumn{1}{c|}{$\bm{f^{(MMSE)}}$} & {$$} & {$$} & {$$} & {$$} & {$$} & {$$} & {$$} & {$$} & {$$} & {$10,646,386$} & \multicolumn{1}{c|}{$10,661,291$} & {$25,541$} \\
\hline \multicolumn{1}{c|}{$\bm{f^{(CL)}}$} & {$3$} & {$$} & {$$} & {$$} & {$$} & {$$} & {$$} & {$$} & {$9,734,574$} & {$9,744,764$} & \multicolumn{1}{c|}{$9,758,606$} & {$34,538$} \\
\multicolumn{1}{c|}{$\bm{f^{(MMSE)}}$} & {$$} & {$$} & {$$} & {$$} & {$$} & {$$} & {$$} & {$$} & {$9,734,765$} & {$9,744,500$} & \multicolumn{1}{c|}{$9,758,143$} & {$34,074$} \\
\hline \multicolumn{1}{c|}{$\bm{f^{(CL)}}$} & {$4$} & {$$} & {$$} & {$$} & {$$} & {$$} & {$$} & {$9,837,277$} & {$9,847,906$} & {$9,858,214$} & \multicolumn{1}{c|}{$9,872,218$} & {$85,302$} \\
\multicolumn{1}{c|}{$\bm{f^{(MMSE)}}$} & {$$} & {$$} & {$$} & {$$} & {$$} & {$$} & {$$} & {$9,835,850$} & {$9,846,669$} & {$9,856,516$} & \multicolumn{1}{c|}{$9,870,315$} & {$83,400$} \\
\hline \multicolumn{1}{c|}{$\bm{f^{(CL)}}$} & {$5$} & {$$} & {$$} & {$$} & {$$} & {$$} & {$10,005,044$} & {$10,056,528$} & {$10,067,393$} & {$10,077,931$} & \multicolumn{1}{c|}{$10,092,247$} & {$156,494$} \\
\multicolumn{1}{c|}{$\bm{f^{(MMSE)}}$} & {$$} & {$$} & {$$} & {$$} & {$$} & {$$} & {$10,005,302$} & {$10,055,329$} & {$10,066,390$} & {$10,076,456$} & \multicolumn{1}{c|}{$10,090,563$} & {$154,811$} \\
\hline \multicolumn{1}{c|}{$\bm{f^{(CL)}}$} & {$6$} & {$$} & {$$} & {$$} & {$$} & {$9,419,776$} & {$9,485,469$} & {$9,534,279$} & {$9,544,580$} & {$9,554,571$} & \multicolumn{1}{c|}{$9,568,143$} & {$286,121$} \\
\multicolumn{1}{c|}{$\bm{f^{(MMSE)}}$} & {$$} & {$$} & {$$} & {$$} & {$$} & {$9,400,832$} & {$9,466,638$} & {$9,513,971$} & {$9,524,436$} & {$9,533,961$} & \multicolumn{1}{c|}{$9,547,308$} & {$265,286$} \\
\hline \multicolumn{1}{c|}{$\bm{f^{(CL)}}$} & {$7$} & {$$} & {$$} & {$$} & {$8,445,057$} & {$8,570,389$} & {$8,630,159$} & {$8,674,568$} & {$8,683,940$} & {$8,693,030$} & \multicolumn{1}{c|}{$8,705,378$} & {$449,167$} \\
\multicolumn{1}{c|}{$\bm{f^{(MMSE)}}$} & {$$} & {$$} & {$$} & {$$} & {$8,437,023$} & {$8,545,017$} & {$8,604,832$} & {$8,647,856$} & {$8,657,369$} & {$8,666,026$} & \multicolumn{1}{c|}{$8,678,159$} & {$421,947$} \\
\hline \multicolumn{1}{c|}{$\bm{f^{(CL)}}$} & {$8$} & {$$} & {$$} & {$8,243,496$} & {$8,432,051$} & {$8,557,190$} & {$8,616,868$} & {$8,661,208$} & {$8,670,566$} & {$8,679,642$} & \multicolumn{1}{c|}{$8,691,971$} & {$1,043,242$} \\
\multicolumn{1}{c|}{$\bm{f^{(MMSE)}}$} & {$$} & {$$} & {$$} & {$8,236,916$} & {$8,417,305$} & {$8,525,046$} & {$8,584,722$} & {$8,627,645$} & {$8,637,136$} & {$8,645,773$} & \multicolumn{1}{c|}{$8,657,877$} & {$1,009,148$} \\
\hline \multicolumn{1}{c|}{$\bm{f^{(CL)}}$} & {$9$} & {$$} & {$8,470,989$} & {$9,129,696$} & {$9,338,521$} & {$9,477,113$} & {$9,543,206$} & {$9,592,313$} & {$9,602,676$} & {$9,612,728$} & \multicolumn{1}{c|}{$9,626,383$} & {$3,950,814$} \\
\multicolumn{1}{c|}{$\bm{f^{(MMSE)}}$} & {$$} & {$$} & {$8,467,380$} & {$9,118,521$} & {$9,318,217$} & {$9,437,490$} & {$9,503,553$} & {$9,551,070$} & {$9,561,577$} & {$9,571,138$} & \multicolumn{1}{c|}{$9,584,538$} & {$3,908,970$} \\
\hline{} & {$\widehat{f}_j^{(CL)}$} & {$1.4925$} & {$1.0778$} & {$1.0229$} & {$1.0148$} & {$1.0070$} & {$1.0051$} & {$1.0011$} & {$1.0010$} & {$1.0014$} & & {$6,047,061$}\\
{} & {$\sigma_j^{(CL)}$} & {$135.253$} & {$33.803$} & {$15.760$} & {$19.847$} & {$9.336$} & {$2.001$} & {$0.823$} & {$0.219$} & {$0.059$} & & {$5,918,083$} \\
{} & {$\widehat{f}_j^{(MMSE)}$} & {$1.4919$} & {$1.0769$} & {$1.0219$} & {$1.0128$} & {$1.0070$} & {$1.0050$} & {$1.0011$} & {$1.0010$} & {$1.0014$} & {$$}\\
{} & {$\sigma_j^{(MMSE)}$} & {$154.221$} & {$33.000$} & {$16.770$} & {$22.397$} & {$8.300$} & {$2.166$} & {$0.720$} & {$0.158$} & {$0.041$} & {$$}\\
\end{tabular}
} }  }
\end{center}
\caption{Predicted cumulative CL claims $\widehat{C}_{i,j}^{(CL)}$
for actual data and estimated CL reserves
$\widehat{C}_{i,J}^{(CL)}-C_{i,J-i}$ under the classical and
Bayesian DFCL models.} \label{tab5}
\end{table}
}

\end{landscape}

\newpage

\begin{landscape}
{\footnotesize
\begin{table}[ptbh]
\begin{center}
{\footnotesize {\scriptsize {\
\begin{tabular}{|c|c|c|c|c|c|c|c|c|c|c|}
\hline
Accident Year $i$ & 1 & 2 & 3 & 4 & 5 & 6 & 7 & 8 & 9 & Total\\
\hline \hline
$\left(C_{i,I-i}\widehat{\Gamma}^{freq}_{I-i}\right)^{1/2}$ & 192 & 740 & 2,668 & 6,831 & 30,474 & 68,207 & 80,071 & 126,952 & 389,768 & 424,361 \\
\hline
$\left(C_{i,I-i}^2\widehat{\triangle}^{freq}_{I-i}\right)^{1/2}$ & 503 & 1,560 & 3,059 & 12,639 & 25,761 & 20,776 & 33,771 & 41,554 & 108,547 & 157,680\\
\hline
$\left(\text{msep}_{C_{i,J}|\mathcal{D}_I}^{freq}\left(\widehat{C}_{i,J}\right)\right)^{1/2}$
& 538 & 1,727 & 4,059 & 14,367 & 39,904 & 71,301 & 86,901 &
133,580 & 404,601 & 452,708 \\ \hline
$Vco_i(\%)$ & 3.61\% & 6.76\% & 11.91\% & 17.02\% & 25.61\% & 25.00\% & 19.38\% & 12.81\% & 9.93\% & 7.49\%\\
\hline \hline
$\left(C_{i,I-i}\widehat{\Gamma}^{Bayes}_{I-i}\right)^{1/2}$ & 134 & 533 & 2,307 & 7,185 & 27,367 & 74,235 & 86,404 & 129,038 & 437,482 & 470,982\\
\hline
$\left(C_{i,I-i}^2\widehat{\triangle}^{Bayes}_{I-i}\right)^{1/2}$ & 224 & 894 & 1,801 & 4,327 & 15,819 & 29,861 & 32,243 & 49,198 & 152,879 & 211,633 \\
\hline
$\left(\text{msep}_{C_{i,J}|\mathcal{D}_I}^{Bayes}\left(\widehat{C}_{i,J}\right)\right)^{1/2}$ & 261 & 1,040 & 2,927 & 8,387 & 31,610 & 80,016 & 92,224 & 138,099 & 463,425 & 504,934\\
\hline
$Vco_i(\%)$ & 1.75\% & 4.07\% & 8.59\% & 10.06\% & 20.42\% & 30.16\% & 21.86\% & 13.68\% & 11.86\% &  8.53\%\\
\hline
$\text{VaR}^{Bayes}_{0.95}\left(C_{i,J}-E[C_{i,J}|\mathcal{D}_I]|\mathcal{D}_I\right)$ & 554 & 2,183 & 5,632 & 15,820 & 61,122 & 152,531 & 173,665 & 161,619 & 816,701 & 910,757\\
\hline
$\text{VaR}^{Bayes}_{0.99}\left(C_{i,J}-E[C_{i,J}|\mathcal{D}_I]|\mathcal{D}_I\right)$ & 726 & 2,918 & 7,430 & 22,515 & 79,472 & 201,322 & 228,448 & 211,125 & 1,278,665 & 1,454,966 \\
\hline \hline
$\left(C_{i,I-i}\widehat{\Gamma}^{cred}_{I-i}\right)^{1/2}$ & 192 & 740 & 2,668 & 6,831 & 30,474 & 68,207 & 80,071 & 126,952 & 389,769 & 424,362\\
\hline
$\left(C_{i,I-i}^2\widehat{\triangle}^{cred}_{I-i}\right)^{1/2}$ & 188 & 534 & 1,493 & 3,391 & 13,515 & 27,284 & 29,674 & 43,901 & 129,764 & 185,015 \\
\hline
$\left(\text{msep}_{C_{i,J}|\mathcal{D}_I}^{cred}\left(\widehat{C}_{i,J}\right)\right)^{1/2}$ & 269 & 913 & 3,057 & 7,627 & 33,337 & 73,462 & 85,392 & 134,329 & 410,802 & 462,941\\
\hline
$Vco_i(\%)$ & 1.81\% & 3.58\% & 8.97\% & 9.04\% & 21.40\% & 25.77\% & 19.04\% & 12.88\% & 10.40\% & 7.82\%\\
\hline
\end{tabular} } }  }
\end{center}
\caption{Comparison of the frequentist's bootstrap
$\text{msep}^{freq}$, the Bayesian MCMC-ABC $\text{msep}^{Bayes}$
and the credibility $\text{msep}^{cred}$. The coefficient of
variation is as defined in W\"uthrich-Merz \cite{Wuthrich}.}
\label{tab6}
\end{table}
}
\end{landscape}

%
%
\newpage


\newpage

%
%
%


\newpage 

\end{document}